\documentclass[sigconf]{acmart}
\AtBeginDocument{
  }

\copyrightyear{2026}
\acmYear{2026}
\setcopyright{cc}
\setcctype{by-nc-nd}
\acmConference[UIST '26]{The 39th Annual ACM Symposium on User Interface Software and Technology}{November 02--05, 2026}{Detroit, MI, USA}
\acmBooktitle{The 39th Annual ACM Symposium on User Interface Software and Technology (UIST '26), November 02--05, 2026, Detroit, MI, USA}
\acmDOI{10.1145/3830398.3830612}
\acmISBN{979-8-4007-2856-3/2026/11}

\usepackage{_macros}
\usepackage{listings}
\usepackage{xcolor}
\usepackage{float}    
\usepackage{stfloats}
\newcommand{\rhl}[1]{\textcolor{black}{#1}}
\newcommand{\shl}[1]{\textcolor{black}{#1}}

\begin{document}
\title{Surprise2Refine: Axis-Centered Exploration-To-Refinement for Agent-Assisted Creative Scaffolding}
\acmSubmissionID{8681}

\author{Yuzhe You}
\authornote{Work done during an internship at Adobe Research.}
\orcid{0009-0004-7830-4239}
\affiliation{
  \institution{University of Waterloo}
  \city{Waterloo}
  \state{Ontario}
  \country{Canada}
}
\email{y28you@uwaterloo.ca}

\author{Gromit Yeuk-Yin Chan}
\orcid{0000-0003-1356-4406}
\affiliation{
  \institution{Adobe Research}
  \city{San Jose}
  \state{California}
  \country{USA}
}
\email{ychan@adobe.com}

\author{Shunan Guo}
\orcid{0000-0001-5355-8399}
\affiliation{
  \institution{Adobe Research}
  \city{San Jose}
  \state{California}
  \country{USA}
}
\email{sguo@adobe.com}

\author{Anlan Zhang}
\orcid{0000-0003-2371-4631}
\affiliation{
  \institution{Adobe Research}
  \city{San Jose}
  \state{California}
  \country{USA}
}
\email{anlanz@adobe.com}

\author{Eunyee Koh}
\orcid{0000-0003-2091-5972}
\affiliation{
  \institution{Adobe Research}
  \city{San Jose}
  \state{California}
  \country{USA}
}
\email{eunyee@adobe.com}

\author{Jian Zhao}
\orcid{0000-0001-5008-4319}
\affiliation{
    \institution{University of Waterloo}
  \city{Waterloo}
  \state{Ontario}
  \country{Canada}
}
\email{jianzhao@uwaterloo.ca}

\author{Tongyu Zhou}
\orcid{0000-0003-4003-518X}
\affiliation{
  \institution{Adobe Research}
  \city{San Jose}
  \state{California}
  \country{USA}
}
\email{tongyuz@adobe.com}

\begin{abstract}
\rhl{Designers require different design spaces across creative stages: broad during exploration, and targeted during refinement. 
Yet existing agent-driven tools assume a fixed or continuously expanding space, leaving designers to manage and navigate it themselves.}
Informed by a formative study with five designers, \rhl{we propose an axis-centered workflow that adaptively broadens and narrows the design space to support structured exploration and refinement.}
We implemented this workflow in \textsc{Surprise2Refine}, \rhl{a prototype that allows users to build and reshape an $n$×$n$ design space through a set of axis-centered interactions as their creative intent evolves.}
A within-subjects study with 14 designers shows that \textsc{Surprise2Refine} enhances users’ sense of control, supports tracking of scaffolding paths, and \rhl{improves the perceived creativity of design outcomes.}
We further distill design insights to guide future agent-assisted tools for creative scaffolding.
\end{abstract}



\begin{teaserfigure}
  \centering
  \vspace{-15pt}
  \includegraphics[width=0.95\textwidth]{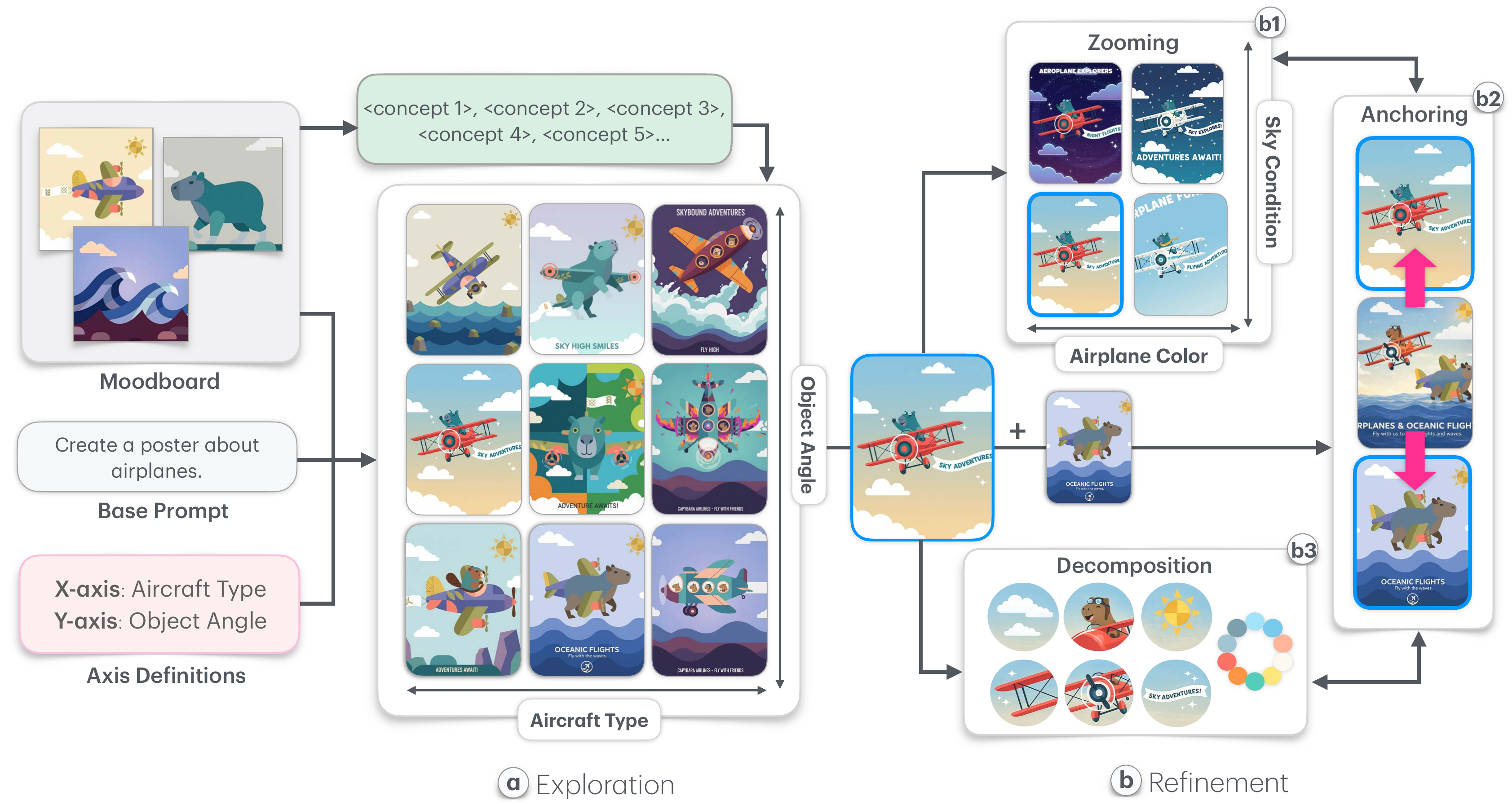}
  \caption{\rhl{Our axis-centered workflow supports adaptive design-space exploration across creative stages.} During early exploration (a), users navigate diverse variations in an $n\times n$ grid along two meaningful conceptual axes. As they move toward refinement (b), zooming (b1), anchoring (b2), and decomposition (b3) reconstruct the space to support increasingly focused, fine-grained adjustments.}
  \Description{A diagram illustrating the Surprise2Refine interaction workflow. On the left, a moodboard with three reference images, a base prompt and axis definitions feed into a 3×3 grid of generated poster designs organized along two axes (e.g., aircraft type and object angle), representing exploration. On the right, the refinement stage shows three interactions: zooming to explore variants of one design along new axes (e.g., sky condition and color), anchoring to select and blend preferred designs into new designs, and decomposition to break a design into reusable elements (e.g., icons, colors). Arrows indicate how users iteratively move from broad exploration to focused refinement.}
  \label{fig:teaser}
\vspace{-4pt}
\end{teaserfigure}

\maketitle

\newcommand{\name}{\textsc{Surprise2Refine}}

\begin{figure}[t]
    \centering
    \includegraphics[width=\linewidth]{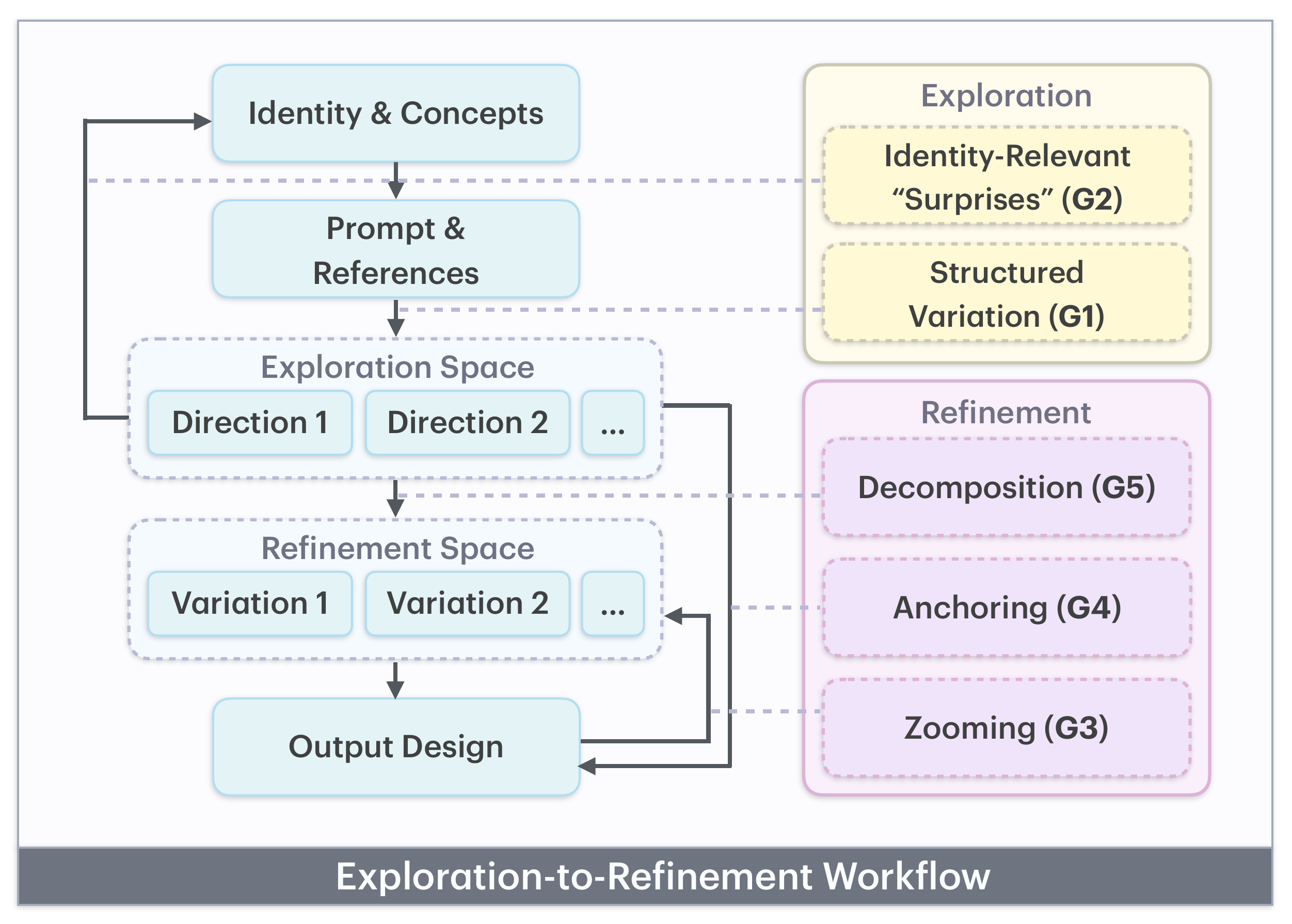}
    \caption{%
    \shl{Overview of designers’ creative scaffolding workflow and how components of our axis-centered workflow supports structured design-space exploration across both stages.}
    }
    \Description{A diagram of an exploration-to-refinement design workflow for creative scaffolding: identity and concepts inform prompts and references, which generate multiple design directions within an exploration space. Selected directions then move into a refinement space containing multiple variations, ultimately producing an output design. Structured variation (G1) supports the transition from prompt and references to the exploration space, and identity-relevant “surprises” (G2) supports the iterative transition from exploration space back to identity and concepts. The workflow then moves into refinement through zooming (G3), anchoring (G4), and decomposition (G5), before producing and iteratively refining the final design. Decomposition (G5) supports the transition from exploration space to finetuning, zooming (G3) supports the transition from the output design back to refinement space, and anchoring (G4) supports the transition from exploration space directly to the output design.}
    \label{fig:workflow}

\end{figure}

\section{Introduction}
\rhl{
French painter Corot once observed, \textit{``I am never in a hurry to reach details. First and above all I am interested in the large masses and the general character of a picture; when these are well established, then I try for subtleties of form and color''}~\cite{goldwater1976artists}.
His account reflects a broader pattern in creative design: with creative freedom, designers often move iteratively from broad exploration toward focused refinement (Fig.~\ref{fig:workflow})~\cite{howard2008describing,botella2018stages,childs2022creativity,abraham2018neuroscience}. 
These changing stages require differently scoped design spaces—i.e., sets of ideas and solutions considered for a task or problem—that expand during divergence and narrow during convergence to match changing information needs~\cite{halskov2021filtering,childs2022creativity,woodbury2006whither}. 
Early on, designers navigate a broad space as they experiment with substantially different styles, colors, and layouts while remaining open to unexpected directions or ``surprises.'' As promising directions emerge, the space becomes more focused, supporting increasingly targeted and fine-grained adjustments.
}

\rhl{
However, existing agent-assisted creative tools provide limited support for design space adaption across exploratory and refinement stages.
Many rely on one-shot generation or iterative prompting \cite{adobe_firefly, canva_ai, microsoft_designer}, where the design space remains implicit and must be inferred through successive outputs. 
Other common approaches such as galleries \cite{adobe_express,leonardo_ai,openai_playground_images, canva_ai, microsoft_designer} make generated alternatives visible, but the outputs are often unstructured or retrospectively organized, while latent-space interfaces \cite{harkonen2020ganspace,odendaal2025dragganspace, liu2019latent} expose a fixed, model-defined space that does not evolve with intent.
While some work has begun to explore adaptive or user-directed design spaces \cite{suh2024luminate, ling2026inspirationgraph, marquardt2025imaginationvellum}, it emphasizes divergence by continually broadening the space—often through infinite canvases—to encourage ideation. 
These approaches offer minimal support for convergence and remain largely user-managed, requiring designers to manually structure and narrow the space as their intent emerges, resulting in excess cognitive load that can hinder creativity~\cite{redifer2019implicit, rodet2022does}.}

To address this gap, we conducted a formative study with five designers to understand how they scaffold creative design, including how they navigate broad and unexpected directions (i.e., ``surprises'') during exploration and transition toward refinement. %
\rhl{We synthesized our insights into an axis-centered workflow for agent-assisted design generation (Fig. \ref{fig:teaser}).}
\rhl{Specifically, designers interact with an adaptable design space represented as an $n \times n$ grid, controlled by two configurable conceptual axes. 
By explicitly defining dimensions, placing design materials along the axes, and invoking stage-specific interactions, semantic scales suited to their intent are constructed: broad and structured for exploring distinct, designer-relevant directions, and smaller, more targeted variations during focused refinement.}
This workflow mirrors the different design spaces required across divergent and convergent stages, supporting open-ended exploration with ``surprises'' in early stages (Fig. \ref{fig:teaser}a) while enabling focused refinement as ideas take shape (Fig. \ref{fig:teaser}b). 

We incorporated this axis-centered workflow into our system, \name{}, which enables designers to first externalize personal intent through a moodboard, then generate diverse candidates in the design space along two meaningful axes. %
The system then provides multiple axis-centered interactions (i.e., zooming (Fig. \ref{fig:teaser}b1), anchoring (Fig. \ref{fig:teaser}b2), decomposition (Fig. \ref{fig:teaser}b3)) to help designers compare, organize, and refine generated candidates within a focused space.
A backend agent pipeline adaptively adjusts generative behavior as users interact, encouraging diversity and ``surprises'' during early exploration and progressively narrowing visual consistency toward the designer's emerging intent during refinement.

We conducted a within-subjects user study with 14 professional designers, comparing \name{} with a baseline version without the axis-centered interactions to understand how the proposed workflow supports designers' scaffolding and helps them make sense of their design decision-making. 
Our results show that participants felt greater control over their design process and reported that \name{} expanded their creative exploration.
Participants also noted that the axis-centered representation helped them better understand both their own design decisions and the agent's generative behavior.
Based on these findings, we distill design insights to inform the development of future agent-assisted design tools that better support creative scaffolding.

To summarize, our contributions in this paper include: 

\begin{itemize}[noitemsep, topsep=0pt]

\item An \rhl{\textbf{axis-centered workflow}}, informed by formative studies, for \rhl{structured design-space exploration}. 
The workflow utilizes a dynamic grid with axes that adapt to intent to support controlled surprises during early exploration and targeted, progressive refinement at later stages.

\item A \textbf{working prototype, \name{}}, that instantiates this workflow through an interface for creative scaffolding, and a backend agent pipeline that adapts its design-space generation strategy across different creative stages.

\item A \textbf{comparative user study} with 14 professional designers that evaluates \name{}'s effectiveness and usability, and surfaces additional insights into how our axis-centered workflow supports designers' creative scaffolding and decision-making.

\end{itemize}

\section{Related Work}

\subsection{Models of the Creative Process}

Prior work has tried to formalize the creative process but varies in its characterization~\cite{howard2008describing,cropley2012psychological,wallas1926art,sadler2015wallas,abraham2018neuroscience, botella2018stages, sadler2015wallas,wallas1926art}.
\shl{
In settings where designers work toward clearly defined goals, some models frame creativity as a problem-solving process, emphasizing structured ideation, analysis, and evaluation in sequences similar to engineering design~\cite{howard2008describing,wallas1926art,lawton2023tool,shneiderman2000creating}.
However, with greater creative freedom, the design process is generally agreed to alternate between divergence and convergence: early stages involve open-ended exploration and ``surprises,'' while later stages focus on increasingly targeted, low-level adjustments.~\cite{abraham2018neuroscience,botella2018stages,cropley2012psychological,childs2022creativity}. 
For instance, the creativity diamond~\cite{childs2022creativity} characterizes this process as expanding toward a broad range of possibilities before narrowing toward selected ideas. Cropley and Cropley~\cite{cropley2012psychological} argue that divergent and convergent thinking are not one-time sequential opposites, but recurring modes that reappear throughout innovation.}
Together, these models point to a scaffolding process in which designers first explore diverse possibilities, remain receptive to unexpected surprises, and only later commit to a direction and focus on refining relevant variations.

\rhl{
In contrast to this divergence-convergence process, existing agent-driven tools largely apply the problem-solving framing, leaving the design space implicit across one-shot or iterative generations~\cite{adobe_express,canva_ai,microsoft_designer,leonardo_ai}.
While some recent tools begin to support adaptive or user-directed design spaces, they address only divergence—one component of the broader scaffolding process—by continually expanding the space, leaving designers to organize emerging directions and manage convergence themselves~\cite{ivanov2022moodcubes,wan2023gancollage,dang2023worldsmith,sarukkai2024block}.
Additionally, these systems often equate divergence with expanding the space as much as possible, without considering whether those dimensions are actually meaningful to designers or aligned with their evolving intent~\cite{suh2024luminate, ling2026inspirationgraph, marquardt2025imaginationvellum}.
Thus, in this work, we explore an axis-centered workflow that supports designers in scaffolding from high-level directions toward increasingly focused refinement while grounding the process in identity-relevant dimensions.
}

\subsection{Agent-Assisted Design Generation}
AI agents are increasingly adopted to support visual design, from generating layouts to selecting assets.
A range of tools reflect this shift, including commericial tools such as Adobe Express \cite{adobe_express}, Microsoft Designer \cite{microsoft_designer}, Canva AI \cite{canva_ai}, and Leonardo.AI \cite{leonardo_ai}, which rely on agent-generated templates and support design through conversational edits.
Other similar tools include Vinci~\cite{guo2021vinci}, which generates advertising posters from product images and taglines; Brickify~\cite{shi2025brickify}, which assembles designs from user-specified tokens and constraints; and GRIDS~\cite{dayama2020grids}, which rapidly generates GUI layout candidates.
\shl{
However, since these tools primarily adopt the problem-solving framing to target applied contexts~\cite{sarukkai2024block,lawton2023tool,shi2025brickify}, their interactions follow a rather linear, sequential pattern: users specify prompts, the system rapidly produces an output, and users revise their prompts to obtain new alternatives.}
\shl{
These tools encourage a process that prioritizes rapid convergence, offering little visibility into the design space and limited support for exploring, structuring, or reshaping it.
}

\shl{
This design orientation is less suited for scaffolding workflows (Fig.~\ref{fig:workflow}) that iteratively move between divergent and convergent stages.}
Although existing agent-assisted tools can generate multiple candidates, outputs are typically treated as isolated responses rather than part of a navigable exploratory space. 
As a result, designers must manually track and organize variations across disconnected generations, making it difficult to maintain a mental model of emerging directions or progressively refine a chosen path.
\rhl{
While some tools adopt galleries that present multiple alternatives in a unified view, these outputs are unstructured or organized retrospectively after generation, rather than produced within a designer-meaningful space that guides generation~\cite{adobe_express,leonardo_ai,openai_playground_images, canva_ai, microsoft_designer}. 
Similarly, latent-space exploration exposes relationships among generated outputs, but the underlying space is fixed and model-defined rather than constructed around the designer’s evolving intent and creative stage~\cite{harkonen2020ganspace,odendaal2025dragganspace, liu2019latent}.}
Our paper addresses this limitation by introducing an axis-centered workflow that enables designers to navigate and track multiple directions within an adaptable design space, then progressively narrow it toward refinement, rather than treating design as a one-pass prompt-to-output translation process.

\subsection{Design Spaces in Human-AI Co-Creation}

\rhl{
Since designers have different information needs across creative stages~\cite{halskov2021filtering}, recent work has begun to explore how agents can help construct design spaces of possibilities.
Luminate~\cite{suh2024luminate} uses semantic axes to expand the design space of possibilities for creative writing, InspirationGraph~\cite{ling2026inspirationgraph} represents evolving visual design alternatives as a continuously growing tree, and ImaginationVellum~\cite{marquardt2025imaginationvellum} uses generative strokes on an infinite canvas to rapidly produce different art variations.}
Other work similarly increases variation or the design space to support early ideation and inspiration: MoodCubes~\cite{ivanov2022moodcubes} and GANCollage~\cite{wan2023gancollage} provide AI-assisted collage space to help users explore visual materials, while WorldSmith~\cite{dang2023worldsmith} and Sarukkai et al.~\cite{sarukkai2024block} generate provisional ideas from multimodal or incomplete inputs.
\rhl{
However, the design spaces in these are either user-managed, or continually expanding with little to no support for convergence. 
For example, Luminate is stated to counter fixation and promote divergent thinking by always enlarging the design space.
Despite acknowledging the importance of convergence, its scenario leaves users to converge on their own by ``selecting several dots randomly~\cite{suh2024luminate}'' and ``manually bookmark[ing]~\cite{suh2024luminate}.''
InspirationGraph~\cite{ling2026inspirationgraph} and ImaginationVellum~\cite{marquardt2025imaginationvellum} do not adapt the space to users’ creative stages or underdeveloped intent, as users have to manually extend a branching tree or spatially guide generation on an open canvas. 
The design spaces in these related systems~\cite{ling2026inspirationgraph,marquardt2025imaginationvellum,ivanov2022moodcubes,wan2023gancollage,dang2023worldsmith,sarukkai2024block} remains user-managed, and assumes that users can clearly articulate the directions they wish to explore.}

Other work has begun to examine ``surprises'' in AI-assisted generation space. 
For example, GANspiration \cite{mozaffari2022ganspiration} aims to generate serendipitous examples by sampling from large design datasets, while DesignAID \cite{cai2023designaid} uses large language models (LLMs) to broaden the conceptual design space for image generation. 
Both aim to increase the likelihood of inspiration by expanding the diversity of generations, equating surprise with diversity. %
However, creativity research characterizes surprise more precisely, describing it as emerging from unexpected idea combinations or emergent insights \cite{boden2004creative}.
Design research further shows that surprise is shaped by designers’ expectations and identities, with designers actively seeking and interpreting outcomes that resonate with their intentions and experiences \cite{abraham2018neuroscience, boden2004creative}.
This suggests that a design space with meaningful surprises arises from the interaction between generated artifacts and a designer’s evolving understanding—shaped by their identity, experiences, and values—rather than from diversity alone.

\rhl{
Taken together, prior work has begun to explore new ways to construct design spaces in human-AI co-creation. 
However, these spaces remain largely user-managed or oriented toward continual divergence, while treating ``surprises'' as a by-product of output diversity.
In contrast, our axis-centered workflow is motivated by evidence that designers require different design spaces across divergence and convergence to match their information needs, as excess cognitive load can hinder creativity~\cite{halskov2021filtering,childs2022creativity,woodbury2006whither,redifer2019implicit, rodet2022does}. 
Accordingly, our approach dynamically restructures the space to support both exploration and refinement. 
It further builds the space around designer-relevant dimensions, enabling ``surprises'' that are not only diverse but also grounded in designers’ identities, supporting more meaningful design outcomes.
}

\section{Formative Study}
\label{Formative Study}

To better understand designers’ scaffolding processes and their challenges, we conducted a formative study with five professional designers (F1$\sim$F5) from diverse backgrounds including visual, UI/UX, and product design. 
The study explored: 
(1) \textit{how designers transition from exploration to refinement};
(2) \textit{how designers handle unexpected outcomes and define “good” or “bad” surprises}; 
and (3) \textit{how they envision agent collaborations across creative stages}.

The study began with a pre-questionnaire and a semi-structured interview to understand their usual design workflows.
Participants then completed a 30-min poster design walkthrough on the theme ``\textit{table tennis}''. 
We intentionally chose only a high-level theme to preserve creative freedom and observe designers' strategies and unique identities under the same topic. 
Participants were encouraged to think aloud, and optional reference materials were provided for examining unexpected surprises with external stimuli.
They were free to use any design tools they normally would.
Finally, a post-interview was conducted to obtain their reflections.

\shl{
Overall, our study confirmed that when given greater creative freedom, designers adopt a creative scaffolding workflow that iteratively shifts between exploration (i.e., divergence) and refinement (i.e., convergence).}
Additionally, the participants are increasingly integrating AI or agent-assisted tools (e.g., MidJourney \cite{midjourney}, GPT \cite{chatgpt}, Nano Banana \cite{nanobanana_model}) into their workflows, but only in problem-solving contexts with clear requirements (F1, F3, F5) or early brainstorming (F2, F4). 
When probed on the characteristics of their AI usage, participants noted that many existing agents appear to be ``\textit{designed from a programmer-oriented perspective}"{F3} and do not align well with their scaffolding process. 
This suggests a gap between designers' openness to using agent-assisted tools and the lack of interaction to support the exploration-to-refinement workflow (Fig.~\ref{fig:workflow}).
\rhl{We used these findings to derive design considerations for structured design-space exploration across creative stages, which directly informed our interaction and system design.}

\subsection{Design Considerations}

The following design considerations are derived to support design space exploration across creative scaffolding (Fig.~\ref{fig:workflow}), from early-stage exploration (\textbf{G1}, \textbf{G2}) to later-stage refinement (\textbf{G3}, \textbf{G4}, \textbf{G5}):

\textbf{G1: Structured diversity across design directions.}
All participants described deliberately exploring substantially different directions during early brainstorming. 
Rather than a single concept, they generate rough designs across diverse compositions, styles, and ideas to conduct early divergent exploration~\cite{childs2022creativity, abraham2018neuroscience, botella2018stages}. 
These early ideas are materialized as quick sketches (F1-3, F5) or AI generations (F2, F4).
This suggests that instead of producing visually similar outputs by repeatedly running the same prompt~\cite{adobe_express, microsoft_designer, canva_ai}, generations should cover distinct regions of the design space to ensure meaningful variation without redundancy.

\textbf{G2: Inclusion of identity-relevant surprises.}
In early exploration, all participants were receptive to external inspirations and may abandon initial directions when a new stimulus resonates with them. 
These pivots often emerge from resonant references (F1-5) or AI generations based on personally meaningful materials (F2-4). 
For example, while browsing ping pong images, F3 encountered a photo of Chinese elders playing that resonated with him, leading him to recall watching the Olympics with older family members and pursue a new design direction.
This suggests that effective surprises are rarely random but are grounded in the user's personal taste, experiences, and creative identity~\cite{abraham2018neuroscience, boden2004creative}, surfaced through their references.
Therefore, systems supporting early exploration should generate surprises grounded in designers’ aesthetic preferences and identity cues, rather than through arbitrary randomness.

\textbf{G3: Controlled refinement by exploring smaller variations.}
All participants noted that once a promising direction emerges, they prefer to refine variations of that design rather than generate entirely new ones, often through iterative, small, controlled changes. 
For example, F2 refined a design of elderly Chinese people playing table tennis and eating hotpots by adjusting facial expressions and gestures. 
Similarly, F3 noted using AI to test targeted adjustments, such as \textit{``different placements of visual assets within a poster''}{F3}, without altering the overall direction. 
This suggests that later-stage refinement should support generating controlled variants that preserve the chosen direction while enabling targeted adjustments.

\textbf{G4: Blending of multiple design directions.}
Participants noted that during refinement, they may not immediately commit to a single direction, as several promising designs can emerge simultaneously. 
In these situations, designers often combine elements from different directions to generate new variants. 
For example, F4 used MidJourney’s blending feature to merge multiple designs and explore how their visual elements could work together. 
This suggests that interactions supporting later-stage refinement should also enable designers to blend selected designs in controlled ways, allowing them to iteratively narrow the design space while deciding which elements to retain or discard.

\textbf{G5: Extraction and recombination of specific design elements.}
Participants also described wanting to combine specific elements from different designs rather than blending entire directions. 
\textit{``I like this part of this design, but I prefer the [element] of that one.''}{F2, F3, \& F4}. 
However, they found it difficult to express these preferences through natural language prompts in existing agent-assisted tools. 
This suggests that refinement interactions should also allow designers to extract and recombine specific elements across designs through direct selection, enabling finer control over how visual components are integrated in subsequent generations.

\begin{figure*}[t]
    \centering
    \includegraphics[width=\linewidth]{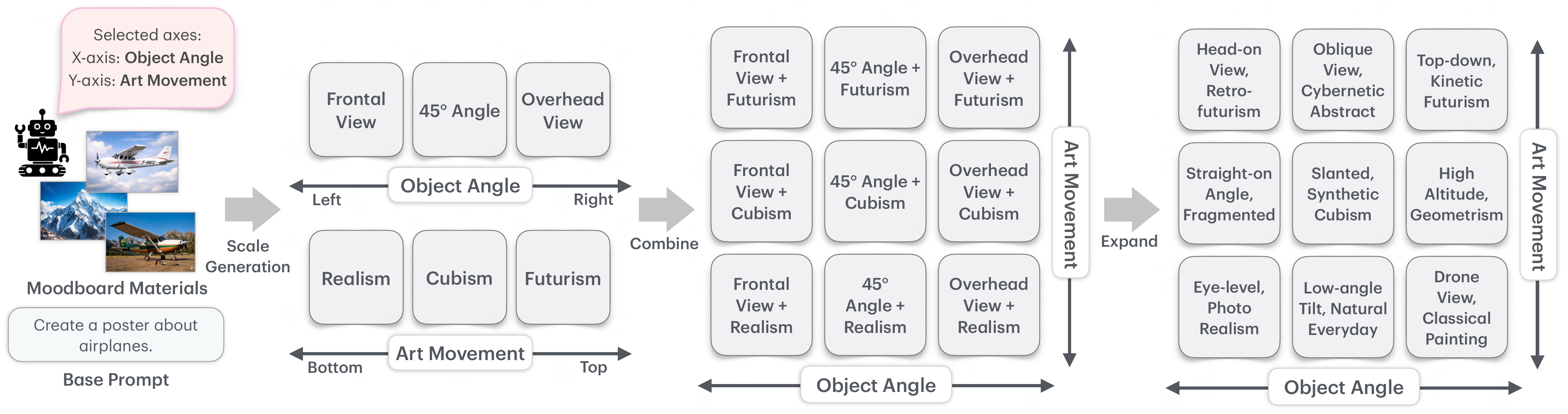}
    \caption{Scale-based agent pipeline for initial generation. Axes, moodboard concepts, and base prompts are used to construct semantic scales along the X and Y axes, which are then combined and expanded to create unique concept pairs for each cell's prompt template in the \(n \times n\) grid.}
    \Description{A diagram showing the scale-based agent pipeline for initial generation. On the left, moodboard images and a base prompt (“Create a poster about airplanes”) are used to define two axes: object angle (e.g., frontal, 45 degree angle, overhead) and art movement (e.g., realism, cubism, futurism). The system first generates semantic scales along each axis, then combines them into a 3×3 grid of concept pairs (e.g., “Frontal View + Futurism,” “45 degree Angle + Cubism”). On the right, these combinations are further expanded into more detailed prompt variations (e.g., “Head-on View, Retro-futurism,” “Slanted, Synthetic Cubism,” “Drone View, Classical Painting”). Arrows indicate the progression from scale generation to combination and expansion across the grid.}
    \label{fig:initial}
\vspace{-15pt}
\end{figure*}

\section{\name{}}

We synthesized these considerations into an axis-centered workflow for design-space exploration. 
Under this workflow, designers interact with an $n$×$n$ grid along two conceptual axes, each representing a design dimension. 
The axes first span large variations, producing controlled yet ``surprising'' outputs that differ significantly across the grid (\textbf{G1}, \textbf{G2}). 
They can then dynamically expand to reduce variation while maintaining a structured search space, allowing users to focus on specific directions of interest (\textbf{G3}).
Designs can be repositioned within the grid to enable interpolation and recombination between neighboring examples, supporting blending and adaptation of design elements within design refinement (\textbf{G4}, \textbf{G5}).

We implemented a working prototype, \name{}, based on this workflow. 
Below, we describe the interactions and backend pipelines of \name{} in detail.
Additional implementation details, including low-level pipeline specifications and prompt templates, are included in Appendix \ref{implementation_details}.

\subsection{Early-stage Exploration (G1, G2)}

During this stage, the system generates diverse variations (\textbf{G1}) consistent with the base prompt and moodboard (Fig. \ref{fig:teaser}a), while introducing controlled “surprises” to inspire new directions (\textbf{G2}).

\subsubsection{Interface Setup and Identity Expression.}

The interface consists of a moodboard and a $n$×$n$ generation grid. 
While our approach generalizes to grids of arbitrary sizes, we use a 3×3 grid to balance design variations with generation time. 
This also maintains cognitive manageability, as prior work suggests that humans can process roughly nine items in short-term memory \cite{miller1956magical}.
The moodboard supports three types of materials to externalize designers' identities, as found by our formative studies (Section~\ref{Formative Study}): \textit{images}, \textit{sketches}, and \textit{text notes}. 
Text entries can be standalone or linked to images or sketches to provide additional context.
The agent, through a vision–language model (VLM) (Gemini-2.5-Flash-Lite), extracts semantic features from the inputs by returning a fixed-cardinality keyword set encoding salient semantic attributes.
These concepts are ranked by the VLM based on their prominence across the extracted features, with the most salient dimensions selected as axes. 
Designers provide a high-level prompt describing their design goal, and can also specify custom axes.
These axes represent conceptual dimensions of variation (e.g., composition, art style) and determine how designs are created and organized within the grid (\textbf{G1}).

\subsubsection{Scales and ``Surprise'' Generation.}
\label{scale generation}

Once a pool of candidate concepts associated with the references is formed (e.g., \textit{vibrant colors}, \textit{cubist abstraction}, \textit{Bauhaus geometry}),
the VLM agent constructs a three-point semantic scale for each axis (Fig. \ref{fig:initial}) by selecting descriptors most relevant to it from the pool. 
This ensures that variations remain consistent with the designer’s identity as inferred from the materials.
Each point on the scale is then expanded into three additional variations, each capturing a different interpretation of the concept (Fig.~\ref{fig:initial}). 
The prompt templates used for this stage are included in Appendix~\ref{initial_gen_appendix}.
These semantic variations are assigned via a row-major, matrix-based mapping, where each cell corresponds to a unique $(\text{X-scale point}, \text{Y-scale point})$ pair. 
Each pair is selected from lexical variants using a fixed modular rule on the cell index, ensuring distinct combinations across cells (\textbf{G1}).

The agent samples 1-3 moodboard materials as visual grounding sources, using a weighting mechanism (Section \ref{sec:natural_decay}), for each axis variation. 
Sketches are assigned higher weights as designers reported stronger ownership and creative intent in them.
The agent expands these inputs into a structured prompt by combining a fixed template—reference instructions, moodboard inputs, and per-cell axis constraints—with the base prompt. 
The user prompt is appended under a ``\textit{Create:}'' header to preserve the overall intent. %
This introduces two forms of controlled ``surprise'' (\textbf{G2}): (1) the semantic scales ensure that each cell explores a distinct region of the design space while remaining aligned with the moodboard and (2) under-specified attributes are treated as flexible variables during prompt expansion. 
These variables are further varied to avoid repetitive outputs.

\begin{figure}[t]
    \centering
    \includegraphics[width=\linewidth]{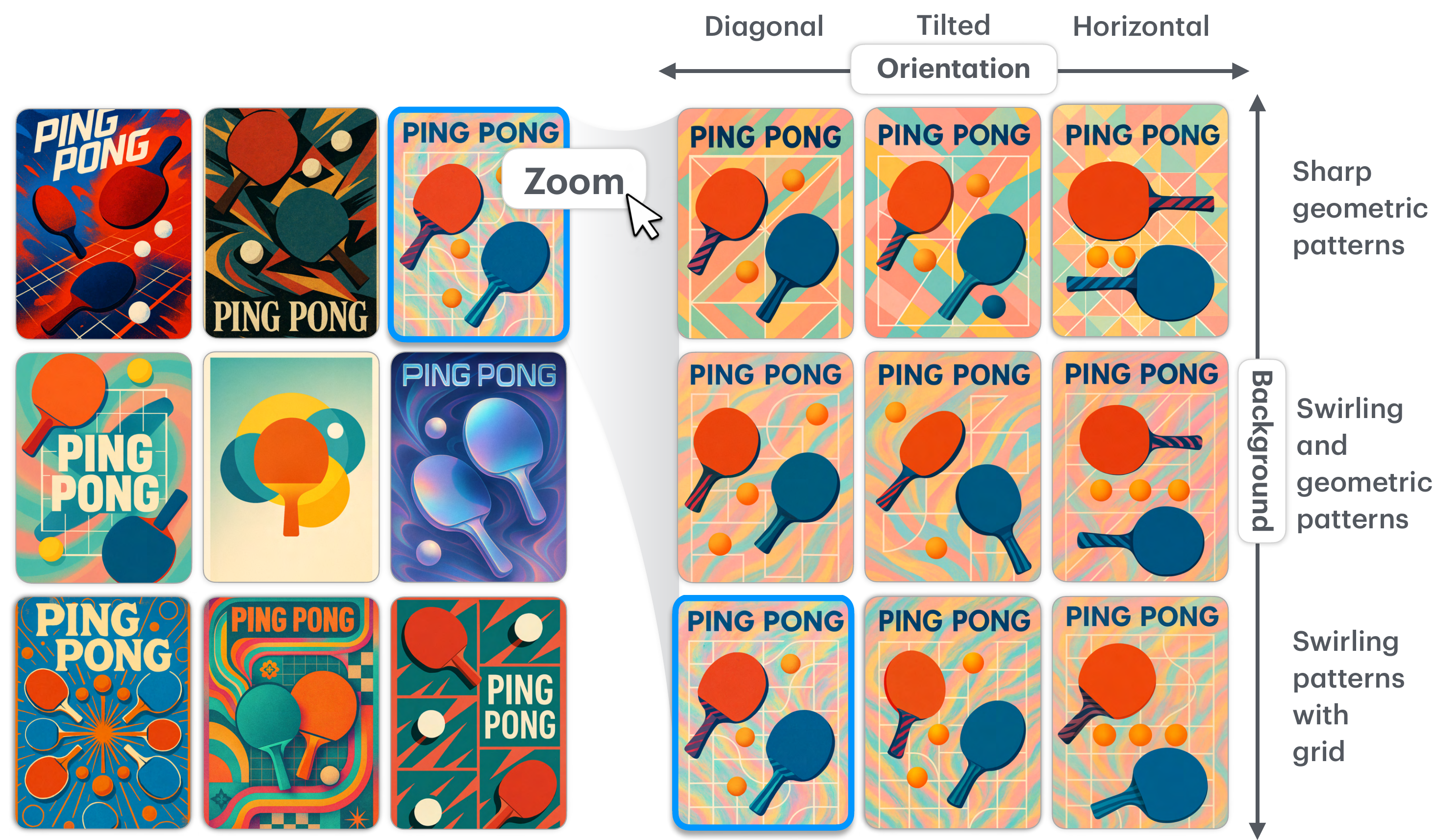}
    \caption{Example of zooming. 
    The axes expand to reduce variation, generating refined variants with small changes along each axis.
    }
    \Description{A diagram illustrating the zooming interaction. On the left, a 3×3 grid of diverse poster designs for “Ping Pong” is shown. A cursor selects one design, triggering a zoom operation. On the right, a new 3×3 grid displays refined variants of the selected design, with reduced variation. The horizontal axis represents orientation (diagonal, tilted, horizontal), and the vertical axis represents background patterns (sharp geometric, swirling geometric, swirling with grid). The refined designs vary subtly along these axes, demonstrating how zooming narrows the design space while maintaining structure.}
    \label{fig:zoom}
\vspace{-15pt}
\end{figure}

\begin{figure}[t]
    \centering
    \includegraphics[width=\linewidth]{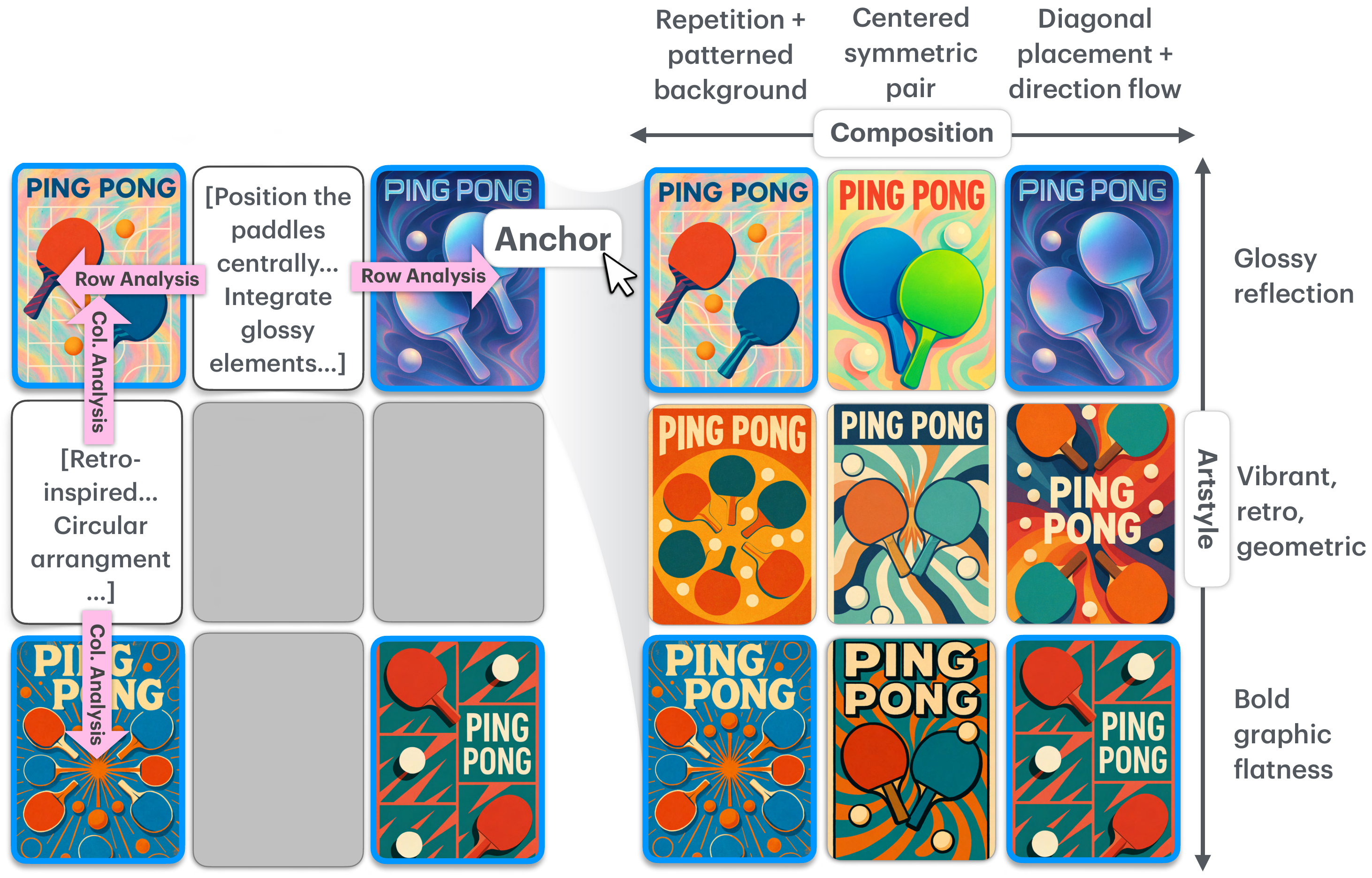}
    \caption{Example of anchoring four posters at the grid corners. The agent extrapolates features from these anchors to create intermediate designs, while maintaining structured variation along the axes.}
    \Description{A diagram illustrating the anchoring interaction. On the left, a partially filled 3×3 grid shows four selected poster designs placed at the corners, with annotations indicating row and column analysis of their features (e.g., composition, style, glossy elements). On the right, a completed 3×3 grid displays interpolated designs generated between the corner anchors. The horizontal axis represents composition (e.g., repetition with patterned background, centered symmetric pair, diagonal placement with directional flow), and the vertical axis represents art style (e.g., glossy reflection, vibrant retro geometric, bold graphic flatness). The intermediate designs blend features from the anchors while maintaining structured variation along both axes.}
    \label{fig:anchor}
\end{figure}

\subsection{Later-stage Refinement (G3, G4, G5)}

To support refinement, the system provides \textit{zooming} (\textbf{G3}), \textit{anchoring} (\textbf{G4}), and \textit{decomposition} (\textbf{G5}) to help users achieve a satisfactory result. 
The workflow also supports progressive convergence via \textit{natural decay} (\textbf{G2, G3}) of contributory moodboard materials.

\subsubsection{Zooming.}
Designers can select any cell to zoom in for the next generation; surrounding cells then become variations of that design rather than entirely new directions (\textbf{G3}).
The VLM agent extracts a concept list describing the zoomed image, then combines these concepts with the grid position, axes, and base prompt to construct updated 3-point scales (Fig. \ref{fig:zoom}), such that the intersection level is aligned to the analyzed image and the other levels are contrastive.
The zoomed image becomes the reference for the new grid, and the agent generates surrounding variants with controlled changes along the axes.
This allows designers to progressively explore variations around a promising design discovered (\textbf{G3}).

The system also modulates the influence of the zoomed image using a zoom-level weighting mechanism. 
Specifically, we define the probability of incorporating additional references as $P_{\mathrm{MB}}(z)=\max(0,\alpha - \beta z)$, where $z$ denotes the zoom level, $\alpha$ is the initial inclusion probability, and $\beta$ controls the decay rate. 
For each cell, a Bernoulli variable determines whether moodboard references are included, with $\mathbb{E}[\mathbf{1}_{\mathrm{MB}}] = P_{\mathrm{MB}}(z)$.
Higher zoom levels increase the influence of the zoomed image during generation, producing variants that remain closer. 
Lower zoom levels allow broader variation, enabling the system to introduce additional contextual changes or incorporate reference cues. 
In this way, zooming enables designers to smoothly transition from broad exploration toward focused refinement while maintaining controlled diversity in the surrounding design space (\textbf{G1}, \textbf{G3}).

\begin{figure}[t]
    \centering
    \includegraphics[width=0.8\linewidth]{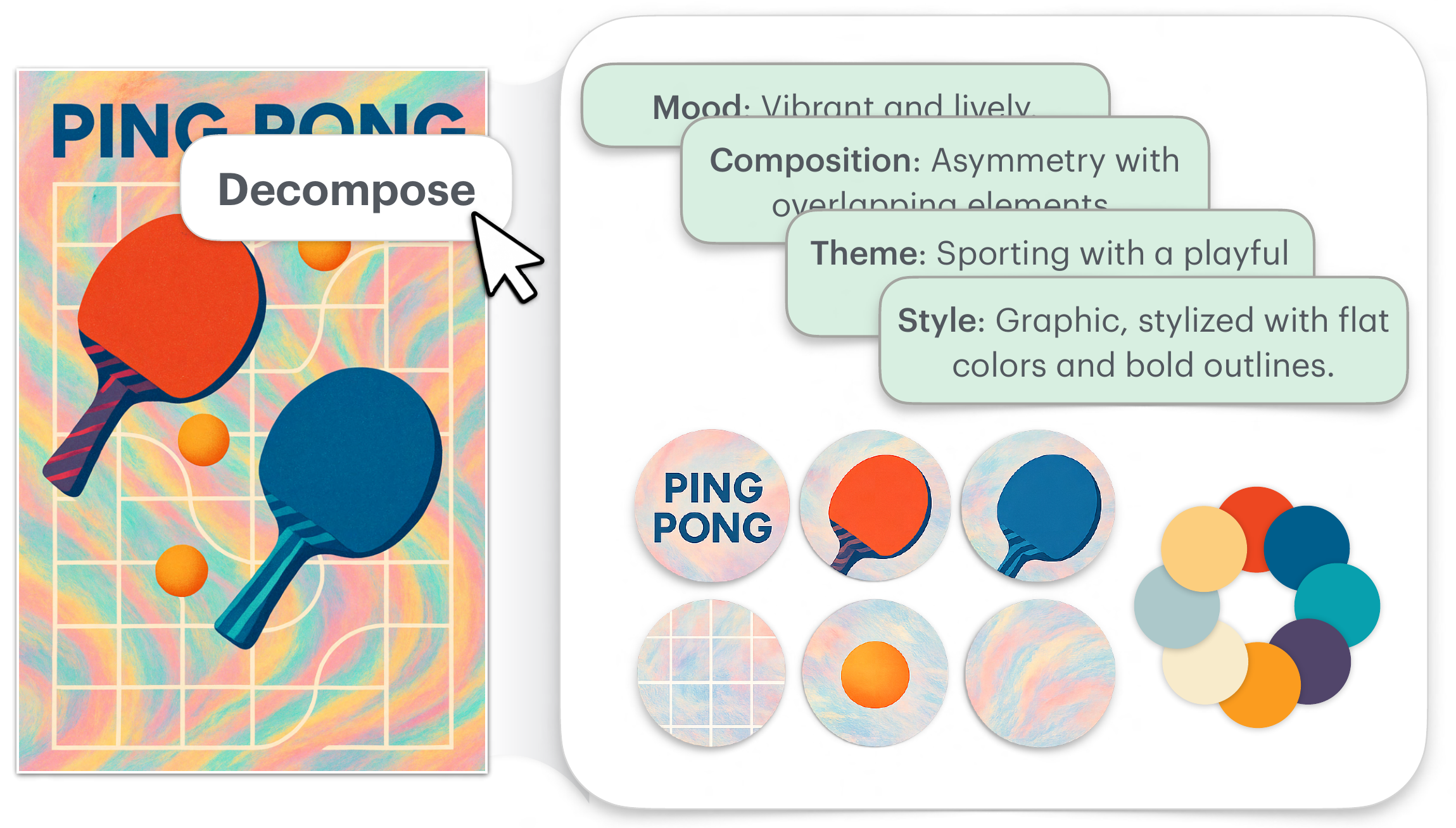}
    \caption{Example of decomposing a poster into textual blocks describing key themes and concepts, along with circular components that extract key visual elements and the color palette.}
    \Description{A diagram illustrating the decomposition interaction. On the left, a “Ping Pong” poster with paddles and a colorful swirling background is selected, with a cursor clicking a “Decompose” button. On the right, the poster is broken down into structured outputs: text blocks summarizing key attributes (e.g., mood, composition, theme, and style), and circular visual elements representing extracted components such as icons, background textures, and a color palette. The layout shows how a single design is analyzed into reusable textual and visual building blocks.}
    \label{fig:decompose}
\end{figure}

\subsubsection{Anchoring.}
Anchoring allows designers to blend visuals from multiple designs (\textbf{G4}) (Fig. \ref{fig:anchor}). 
Designers can drag images from the moodboard into up to  n×n cells or use the ``\textit{keep}'' function to retain generated results directly in the grid. 
The agent then analyzes the anchored images to extract concise multimodal design concepts, returned as a bounded set of short keywords, describing their visual characteristics. 
The extracted keywords are then used to construct the grid’s semantic structure: each anchored cell $(r, c)$ defines its corresponding X and Y axis levels, with these semantics propagated across the row and column to maintain consistency. 

The agent generates intermediate designs by interpolating between anchors. 
This interpolation is spatial: %
cells sharing a row/column with an anchor receive stronger semantic influence, while cells at their intersections blend characteristics from multiple anchors.
The grid is progressively populated, starting from cells with stronger anchor influence, before expanding to less-constrained cells.
The system maintains an internal grid state that tracks filled and unfilled cells. 
Newly generated designs become additional references for subsequent generations, allowing the grid to gradually converge toward coherent design directions while preserving diversity across cells.
This allows designers to explore blended design directions that combine visuals from multiple sources while preserving structured variation across the design space (\textbf{G1, G4}).

\subsubsection{Decomposition.}
\label{sec:decompose}

Designers can extract reusable components from moodboard materials to support element-level recombination (\textbf{G5}). 
This is achieved through: (1) vision–language analysis to extract structured, more abstract visual factors (e.g., theme, style), (2) reference-conditioned image generation to synthesize a unified composite representation, and (3) grid slicing to decompose the composite into individual, reusable design tokens. 
The VLM-based agent first generates holistic descriptions including the overall theme, composition, style, and mood. 
Then, it extracts visual concept elements representing salient components of the image. These are displayed as draggable visual tokens (circles), each representing a distinct concept such as a visual motif, compositional element, or stylistic feature. 
The agent also extracts the dominant color palette of the image as a separate palette token. 
Designers can drag reusable tokens, such as visual concepts, stylistic elements, or color palettes, onto existing grid designs to influence subsequent generations. 
By allowing selective combinations of elements from different sources, decomposition supports fine-grained control over how specific visual features are integrated into new designs (\textbf{G5}).

\begin{figure*}[t]
    \centering
    \includegraphics[width=\linewidth]{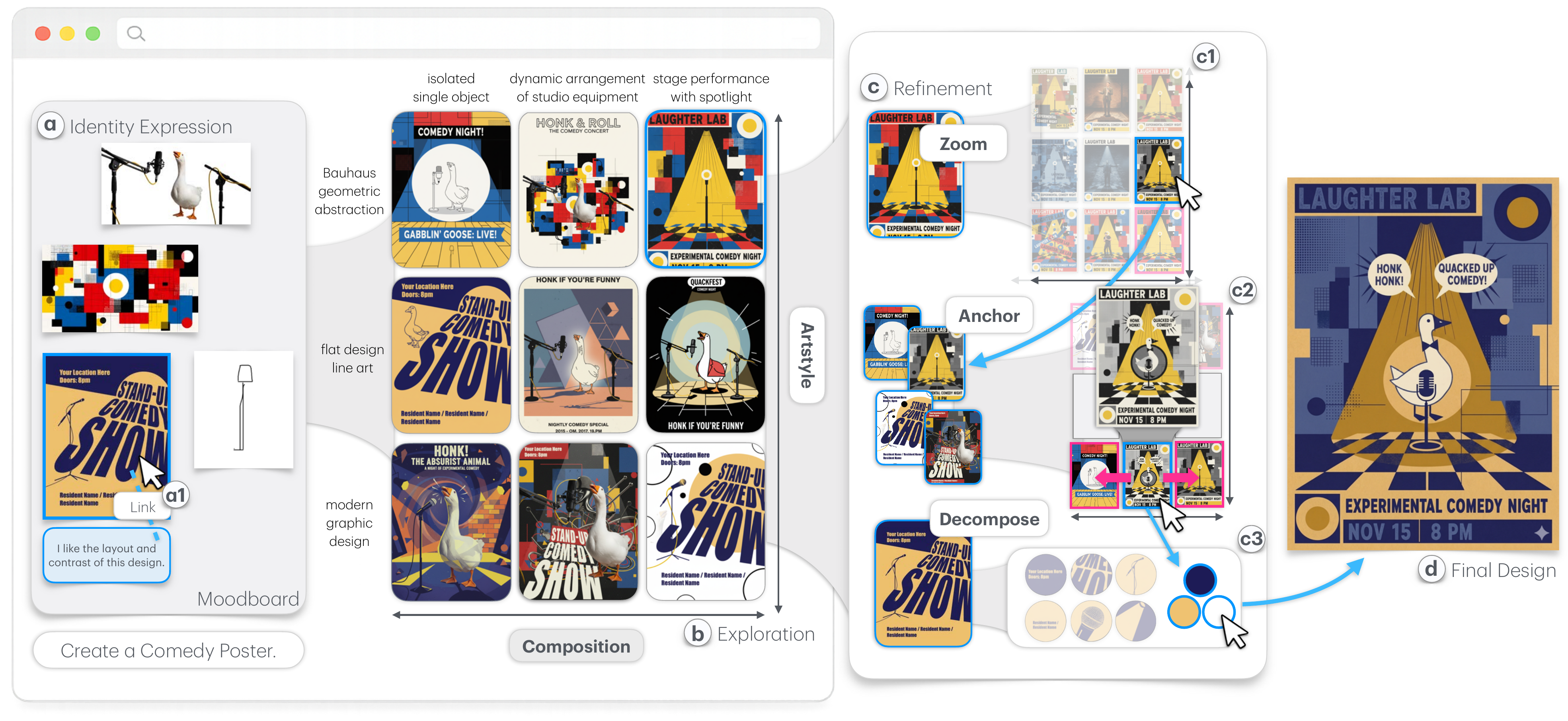}
    \caption{Example of a designer using \name{} to create a poster for a self-funded comedy show.
    The designer begins by adding references she resonates with to the moodboard (a), which generate diverse design variations in the 3×3 grid (b), and then iteratively uses zooming (c1), anchoring (c2), and decomposition (c3) to refine and blend elements into a final poster she is satisfied with (d).}
    \Description{A multi-stage diagram showing a designer using Surprise2Refine to create a comedy poster. On the left (a), the designer adds images and notes to a moodboard alongside a base prompt (“Create a Comedy Poster”). She links a text block that says "I like the layout and contrast of this design" to an existing image. In the middle (b), the system generates a 3×3 grid of diverse poster designs organized along axes such as composition and art style, representing exploration. On the right (c), the designer refines the results using three interactions: zooming (c1) into a selected design to generate closer variations, anchoring (c2) multiple designs to blend their features, and decomposition (c3) to extract reusable elements like colors and components. Arrows indicate the iterative refinement process. On the far right (d), a final poster is shown, featuring a stylized goose under a spotlight with bold geometric design elements and event details.}
    \label{fig:scenario}
\end{figure*}

\subsubsection{Natural Decay.}
\label{sec:natural_decay}
\name{} also supports gradual convergence of designs by implicitly adapting generation preferences based on interactions (\textbf{G2}, \textbf{G3}). 
Specifically, it dynamically adjusts the influence of reference materials according to the designer’s evolving preferences.
It works in conjunction with the probabilistic selection of moodboard materials, where higher-weighted materials are more likely to be chosen as generation references. Given the current weight $w_s^{(t)}$ of a reference image, its weight for the next generation is $\mathrm{clip}( \gamma\, w_s^{(t)} + \eta\, \mathbf{1}_{\mathrm{fb}}(s), \; w_{\min},\; w_{\max})$,
where $\mathbf{1}_{\mathrm{fb}}(s)\in\{0,1\}$ indicates positive feedback, $\gamma=0.95$ is the decay factor, $\eta=0.5$ the reinforcement increment, and $w_{\min}=0.1$, $w_{\max}=5.0$ bound the weights; $\mathrm{clip}(x,a,b)=\min(\max(x,a),b)$.
All materials are initialized with equal weight (1.0) to encourage exploration. 
The parameters were empirically chosen via pilot experimentation to produce stable yet responsive updates that felt natural during iterative design.
As designers interact with generations, their actions are interpreted as preference signals: materials associated with selected designs are reinforced, increasing their reuse likelihood. All materials undergo temporal decay after each generation cycle, reducing the influence of less relevant references. 
This enables prioritization of user-preferred materials while maintaining diversity and preventing long-term dominance.

Together, reinforcement and temporal decay implicitly learn the designer’s evolving identity cues during the design process (\textbf{G2}). 
As designers repeatedly select preferred results, generation progressively shifts toward those directions, supporting a natural transition from divergent exploration to convergent refinement without additional explicit interactions (\textbf{G3}).

\section{Usage Scenario}

This usage scenario illustrates how \name{} supports structured design-space exploration in creative scaffolding. 
Elena is designing a poster for her comedy show, a self-driven project with full creative freedom.
She first uploads references to the moodboard (Fig. \ref{fig:scenario}a), including images that capture her preferred aesthetics and posters she finds appealing.
She adds a sketch capturing one layout idea, and a short text annotation linked to a reference poster, stating ``\textit{I like the layout and contrast of this design}'' (Fig. \ref{fig:scenario}a1).

The agent analyzes these materials to extract structured visual concepts, e.g., ``\textit{geometric abstraction, primary colors, Bauhaus influence, ...}''.
Elena enters ``Create a comedy poster'' and clicks ``Generate.'' 
Based on the inputs, the agent selects ``composition'' and ``art style'' as axes to capture high-variation dimensions, and constructs discrete three-point scales along them (Fig. \ref{fig:scenario}b). 
The composition axis includes ``\textit{isolated single object} → \textit{dynamic arrangement of studio equipment} → and \textit{stage performance with spotlight}'', while the art style axis includes ``\textit{Bauhaus geometric abstraction} → \textit{flat design line art} →\textit{modern graphic design}.''
Using these axes, the system generates a 3×3 grid of designs that systematically explore variations while remaining grounded in Elena’s inputs. 
As Elena reviews the results, she is surprised by one design featuring a Bauhaus-style microphone centered on stage and is drawn to this direction.

To explore this direction further, she updates the X-axis to ``\textit{poster layout}'' and the Y-axis to ``\textit{color palette},'' then clicks ``Zoom'' on the design (Fig. \ref{fig:scenario}c1).
The grid restructures accordingly: the color palette axis updates to ``\textit{primary colors} → \textit{monochromatic schemes} → \textit{muted earth tones},'' while the layout axis spans ``\textit{overlapping elements}  →  \textit{photorealistic scenes}  →  and \textit{geometric abstraction}.'' 
Among these variants, Elena identifies a monochromatic design featuring a black-and-white microphone with a yellow spotlight, which she finds particularly appealing, and drags it back to the moodboard. 
She then zooms out to the original level, and anchors this design along with three other posters she resonates with to the corners of the 3×3 grid.
The agent analyzes these anchor designs and interpolates between them, producing new designs that blend and recombine their stylistic and structural elements (Fig. \ref{fig:scenario}c2).
Among these generations, one design catches her attention.
It maintains the monochromatic palette of the microphone poster she selected during zooming, while integrating abstract shapes of a small goose performing a comedy act—drawn from other designs she previously anchored. 
The layout also resembles the reference poster she had earlier noted for its composition. 
This combination of elements aligns well with her intent, but she is interested in exploring alternative color palettes.
To do so, Elena selects another poster whose color palette she prefers and activates ``Decomposition.'' 
The agent decomposes the poster into reusable design tokens and a color palette encoded as a ring of colors (Fig. \ref{fig:scenario}c3). 
Elena drags the extracted palette onto the current design. 
The regeneration preserves the composition of the selected design while adopting the new color scheme, producing a result that closely matches her expectations (Fig. \ref{fig:scenario}d).

\section{Evaluation}
\label{sec:evaluation}

We conducted a within-subjects study comparing \name{} to a baseline to evaluate its support for creative scaffolding and how its axis-based interaction improves design workflows.
We recruited 14 designers (P1$\sim$P14; age 24$\sim$47; eight men, six women) across visual, product, and packaging design, as well as illustration.
Their work spans branding, poster design, digital products, concept art, and motion graphics. 
They were recruited through online postings and word-of-mouth via design communities and professional networks. 
Each participant received \$30 for their time.

We conducted a task-based study to evaluate the usability and impact of \name{}, focusing on three key questions:
\textbf{RQ1}: \textit{How do designers utilize the axis-centered workflow to support their scaffolding?}
\textbf{RQ2}: \textit{In what ways do axis-centered interactions enhance designers’ existing workflows?}
\textbf{RQ3}: \textit{In what ways does the axis-centered workflow support designers’ sensemaking of their decision-making?}
Each session began with a pre-questionnaire collecting demographic information, design background, and descriptions of their existing workflows. 
This was followed by a brief pre-interview to further understand their scaffolding strategies and prior experience with AI-assisted tools.
Participants then completed the main task, interacting with two AI-assisted design tools in a within-subjects setting: \name{} and a baseline version without axis-centered interactions.
\rhl{This self-built baseline was chosen to ensure experimental control, as comparisons with commercial tools (e.g., ~\cite{adobe_express,microsoft_designer,canva_ai,leonardo_ai}) introduce confounds such as model differences and extra interface features. 
Because our contribution is the axis-centered workflow, the most relevant comparison is an otherwise equivalent system without it, allowing us to isolate its effects.}

For each condition, participants were given 20 minutes to design a poster based on a selected topic from a predefined set. 
They used the same topic across both conditions and were encouraged to explore any direction to fully express their ideas. 
\rhl{Using the same topic controlled for differences in the inherent scope and constraints of each topic’s design space, e.g., ``Coke'' is more constrained from brand identity than ``Outer Space.''
Assigning different topics would affect both diversity metrics and perceived outcomes, making it difficult to attribute differences to the tools rather than the subject matter.}
The order of conditions was counterbalanced across participants to mitigate ordering effects.
They began with the same set of three to four references (either provided or self-selected). 
After the initial generation, they were free to explore or refine designs. 
Participants were asked to think aloud, and we recorded screen interactions and system logs to capture their workflows.
After each condition, participants completed a post-questionnaire assessing overall experience, support for scaffolding workflows, perceived workflow improvements, and sensemaking of decision-making, along with the Creativity Support Index (CSI). 
The study concluded with a semi-structured post-interview to gather further insights based on observed behaviors and participant reflections. 
Each session lasted approximately 90 minutes.

\begin{figure*}[t]
    \centering
    \includegraphics[width=0.95\linewidth]{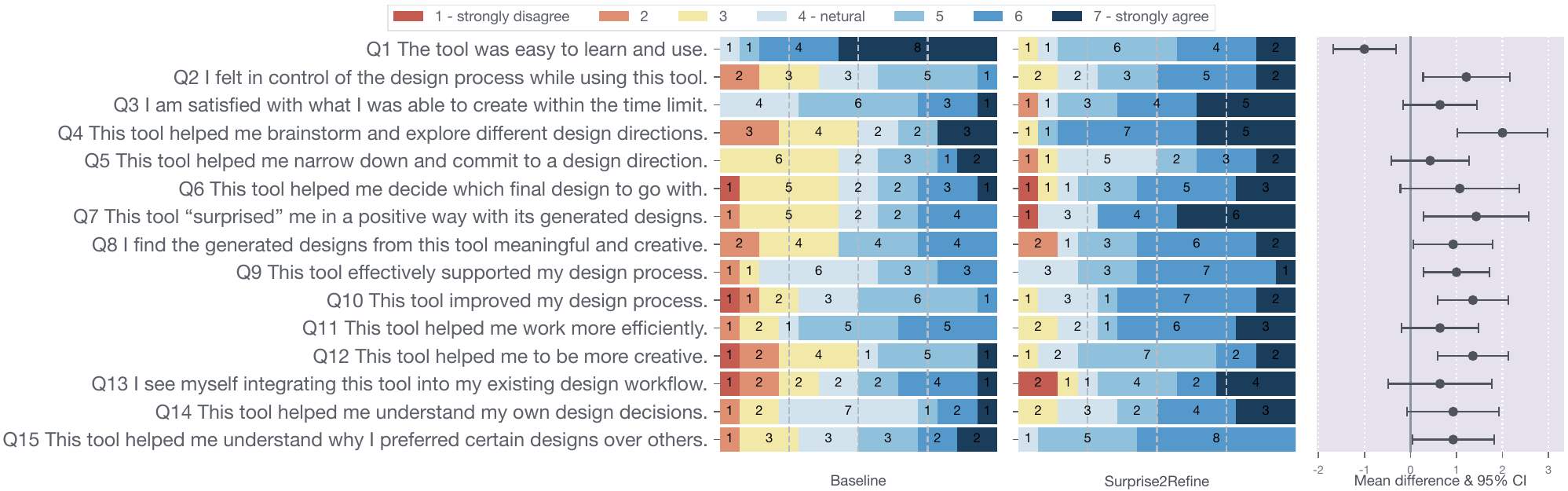}
    \caption{Participants’ post-questionnaire responses on 7-point Likert scales for the baseline and \name{}. Dots indicate mean differences between the two conditions.}
    \Description{A chart showing participants’ responses to 15 questionnaire items (Q1–Q15) on a 7-point Likert scale for two conditions: Baseline and Surprise2Refine. Each row corresponds to a question about usability, control, creativity, and workflow support. For each condition, stacked horizontal bars display the distribution of responses from “strongly disagree” (1) to “strongly agree” (7). On the right, a dot-and-error-bar plot shows the mean difference between conditions with 95 percent confidence intervals for each question. All but one item show higher agreement scores for Surprise2Refine compared to the baseline, indicating improved user experience across measures such as creativity, control, and support for exploration and decision-making. The baseline is rated more positively for being "easy to learn and use."}
    \label{fig:questionnaire}
\vspace{-15pt}
\end{figure*}  
\begin{figure}[t]
    \centering
    \includegraphics[width=\linewidth]{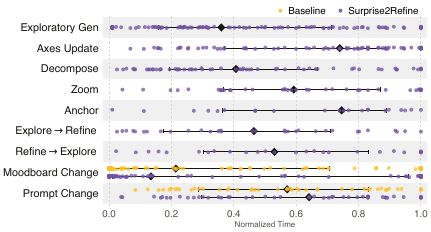}
    \caption{\rhl{Interaction events from all participants across normalized time. 
    While the distribution shows a general progression from exploration to refinement, it also reveals iterative transitions back to exploration.}}
    \Description{A plot showing interaction events from all participants over normalized time. Rows represent exploratory generation, axis updates, decomposition, zooming, anchoring, transitions from exploration to refinement and from refinement back to exploration, moodboard changes, and prompt changes. Purple points represent Surprise2Refine, while yellow points represent the baseline. Black diamonds indicate the median timing of each interaction event, and horizontal whiskers show the interquartile range. The distribution shows a general progression from exploration to refinement: exploratory generations have relatively earlier median timings, whereas refinement interactions have later median timings. However, it also shows that, with Surprise2Refine, designers iteratively transition from refinement back to exploration throughout the design session.}
    \label{fig:timeline}
\end{figure}

\section{Results}
Fig.~\ref{fig:questionnaire} presents participants’ responses from the post-questionnaire.
Additional diversity score calculations and CSI ratings are reported in Appendix \ref{additional_results}.
We used a Wilcoxon signed-rank test to assess differences between conditions, and derived design insights into how \name{} supports and improves creative scaffolding, and aids designers in making sense of their decision-making.

\subsection{RQ1: Support for Scaffolding}

\textbf{Controlled exploration and refinement.} Overall, participants rated \name{} more positively than the baseline across all stages of the creative scaffolding, including exploration (Q4: MD{=6>3.5}; $p{=0.004}$), refinement (Q5:MD{=4.5>4}; $p{=0.329}$), and final design (Q6: MD{=6>4}; $p{=0.043}$). 
Its overall CSI score was also higher (CSI=69.71>55.79; $p{=0.0085}$).
Designers reported that axis-centered interactions gave them greater control (Q2: MD{=5.5 > 4}; $p{=0.019}$), allowing them to anticipate and direct design variations via axes (P2-3, P8-11, P14).
\rhl{Interaction logs also showed that all 14 designers engaged with both exploration-focused (MD{=6 times}) and refinement-focused (MD{=9 times}) axis-centered interactions, and 11 of the 14 designers manually updated their axes during the task (MD{=4.5 times}).}
For example, P14 first defined the Y-axis as ``\textit{art style}'' to explore general styles for a winter-themed poster. 
After noticing an oil painting design, they redefined the Y-axis as ``\textit{oil painting}'' to pursue that direction.
In contrast, the baseline, despite generating multiple outputs, was often perceived as ``\textit{scattered and lacking clear progression}''{P8-9, P14}.
As a result, designers (P3, P8-9, P14) repeatedly added, removed, or replaced references to indirectly influence generation through trial and error. 
This suggests increased cognitive overhead and greater difficulty to track exploration paths, with the system largely determining subsequent outputs.
By making design dimensions explicit and structured via axes, designers could better define and anticipate agent behavior to align generation with their mental model, allowing them to control what to explore/refine next, thus crafting their own design paths.

\textbf{Structured and meaningful surprises.}
\name{} was also rated more positively in how surprising the outputs were (Q7: MD{=6>4}; $p{=0.028}$). 
Mean diversity scores also show that with the same inputs, \name{} produces more diverse generations (Mean LPIPS=0.641>0.529, $p{=0.0001}$).
Though both systems rely on user-provided references, participants (P2–3, P8–11, P14) reported that our tool produced more meaningful surprises by ``\textit{reinterpreting [their] inputs in unexpected yet relevant ways}''{P8}.
For example, when P3 uploaded circuit board images for an exercise poster, the agent surprised them by reinterpreting their pathways as flows of energy, transforming circuit traces into motion lines.
This shows that during early exploration, designers value novel reinterpretations of their materials rather than exact one-to-one mappings.
In addition, participants noted that the axes further amplified the quality of surprises (P2-3, P8-9). 
By organizing variations along explicit dimensions, unexpected results become more interpretable and aligned with designers’ intended directions.
Interestingly, though P8 and P9 also changed their design direction during baseline interaction, they still reported that \name{} provided more positive surprises, as its variations appeared purposeful and coherent. 
While the baseline occasionally produced unexpected outputs, these were often perceived as ``\textit{random hallucinations}''{P8} due to their lack of structure and progression.
P3, P8, and P9 noted that the structured axes helped them ``\textit{break [their] thinking impasse}"{P9} by extending their mental design space along these directions—``\textit{opening up}"{P8} new possibilities—leading to greater creativity (Q12: MD{=5>3.5}; $p{=0.006}$) and broader exploration compared to the ``\textit{`random' feel of the [baseline]}"{P8}.

\textbf{Tracking and revisiting scaffolding paths.}
Participants noted that \name{} supports better tracking of their scaffolding, helping them maintain awareness of their design trajectory. 
P3 stated that the grid-based interaction ``\textit{felt like a map of [their] design journey}" and P2 compared it to ``\textit{version control.}" 
This allows designers to freely explore and refine multiple directions without concern for getting lost, as they could easily return to earlier states.
This was observed across multiple participants (P2–3, P7, P10–11, P14), who frequently experimented with alternative directions before returning to a previous zoom level. 
In many cases, participants only recognized a preferred direction after comparing it against other explored variations, suggesting that maintaining access to prior scaffolding paths is critical for reflective decision-making. 
The axis-based structure, combined with zooming and anchoring, preserves these intermediate states, enabling designers to iteratively branch, revisit, and refine ideas without losing context.
``\textit{I am not worried about losing track because I can always zoom back out.}"{P14}.
In contrast, the baseline made it difficult to track and revisit prior paths, as generated outputs were not organized within a persistent structure. 
Participants reported losing promising directions after exploring alternatives, requiring them to reconstruct previous states or abandon them altogether (P2-3, P10-11).

\subsection{RQ2: Improvement of Workflows}
\textbf{Iterative switching between exploration and refinement.}
\name{} was rated as improving workflows (Q10: MD{=6>4.5}; $p{=0.005}$) by enabling more iterative transitions between exploration and refinement, which participants partly attributed to the easy tracking of scaffolding paths.
This switching differs from context-dependent shifts between problem-solving and creative scaffolding models; instead, it occurs within the scaffolding process between exploration and refinement.
Usually, iteration occurs within each stage rather than across them, as designers may be less likely to return to exploration after refinement due to commitment cost. 
But with \name{}, participants reported being more willing to revisit exploration even after refining a direction (P2-3, P8-11, P13-14).
\rhl{Logs support this pattern as well (Fig.~\ref{fig:timeline}): designers typically followed the scaffolding workflow by progressing from exploration (MD point=36\%) to refinement (MD point=52.8\%). 
However, 10 of them later returned to exploration (MD{=2 times}) at positions spread across the session (MD point=53\%, IQR{=52.9\%}).}
For example, when designing a space-themed poster, P9 was refining a symmetrical layout with planets on each side. 
However, the decomposed notes highlighted an ``\textit{asymmetric composition with a sense of depth and vastness},'' prompting them to incorporate this to explore a new, asymmetric direction, effectively returning to exploration.
This shows that our interaction makes the two stages more iterative, allowing designers to more confidently explore diverse directions while maintaining continuity in their process.
As a result, participants were also willing to spend more time in the exploration stage. 
Several participants encountered satisfactory designs mid-process, but chose to continue exploring anyway (P2–3, P8–11, P13-14).
However, this increased exploration also introduced trade-offs. 
Designers with clearer initial goals tended to prioritize efficiency and, in some cases, preferred the baseline’s more direct, linear workflow. These participants instead favored quickly converging on a desired outcome (P4-6, P12).
P5 noted that the continuous stream of variations made it difficult to determine when to stop. 
This sense of open-ended exploration led to uncertainty about whether a better option might still exist, resulting in potential cognitive load and mental fatigue. 
As a result, the baseline was perceived as easier to use due to its more straightforward, goal-oriented interaction (Q1: MD{=5 < 7}; $p{=0.013}$).

\textbf{Shifting from commander to co-creator.}
When using \name{}, participants perceive themselves less as commanders of the system and more as co-creators beside the agent. 
In the baseline, designers felt responsible for specifying every aspect of generation, with P14 noting that the AI ``\textit{just follows what [they] type.}"
This required designers to invest substantial effort in determining both the next direction and how to articulate it in text.
In contrast, \name{} enabled a more collaborative dynamic. 
By introducing structured reinterpretations, \name{} contributed ideas beyond direct user specification, which allows designers to focus on higher-level decisions such as defining which axes to explore next. 
This creates a bi-directional human–agent workflow in which designers and the agent iteratively inform each other. 
Designers express their intent and identity via axes, prompts and moodboard materials. 
In turn, the agent reinterprets these inputs to generate novel and surprising variations to inspire new directions. 

\subsection{RQ3: Sensemaking of Design Decisions}
\textbf{Design preferences via structured dimensions.}
Participants reported that the interaction helped them better understand their design decisions (Q14: MD{=5.5 > 4}; $p{=0.056}$) and preferences (Q15: MD{=6 > 4.5}; $p{=0.048}$). 
By organizing along two explicit dimensions, designers could more easily attribute their preferences to specific factors such as color palettes, composition, or conceptual themes (P3, P5). 
P5 noted that separating variations across two axes allowed them to observe incremental changes and identify which aspects contributed to its appeal. Similarly, P2 described the grid-based interaction as structuring their mental model, allowing them to group and compare designs by style or composition and connect them to ideas they already had. 
This shows that axis-centered interaction supports sensemaking by making design variations explicit and interpretable, helping designers reflect both on their preferences and plan subsequent actions.
Beyond self-understanding, participants reported that it also helped them better understand the agent’s design decisions. 
By exposing how designs vary along dimensions, the system provided insights into how the agent generates and transforms outputs. P2 noted that this made it easier to ``\textit{learn how to control [their] collaborator}", while P8 reported using the axes to better understand how to communicate with the agent despite the difficulty of interpreting AI generation processes.

\textbf{Specification vs. direction-based thinking.}
We observed that both conditions require cognitive effort, but involve different types of sensemaking. 
In the baseline, designers often engaged in \textit{specification-driven sensemaking}, where they had to explicitly formulate their design intentions and translate them into precise textual prompts (P3-4, P6, P12, P10). 
This process emphasizes preemptively defining the final design and encoding it into prompts. 
For example, P12 first constructed a mental image of the final Coke poster, then iteratively refined the base prompt by specifying ``many bubbles floating,'' and a ``large vertical logo on the left'' to guide the AI toward that target design.
In contrast, \name{} encourages \textit{direction-driven sensemaking}, where designers focus on deciding what to explore next rather than specifying exact outcomes. 
P14, while creating a winter poster with a campfire, first generated multiple design directions with default axes, then reflected on them to identify promising ways to improve the design, ultimately introducing ``\textit{light source} ''as a new axis.
However, this shift and its workload were experienced differently across participants. 
While P10 found \name{} reduced the need for upfront thinking by allowing them to try directions more freely, P14 said it required more deliberate consideration of which directions were worth pursuing.
This shows that axis-centered interaction redistributes sensemaking from low-level specification toward higher-level decision-making, which allows designers to think more about exploration strategies than precise output descriptions.

\section{Discussion}

Here, we reflect on the design of \name{} and discuss key lessons learned, along with implications for future research.

\textbf{Balancing exploration and refinement across designers.}
\shl{While designers acknowledged moving from exploration to refinement during creative scaffolding \cite{howard2008describing,abraham2018neuroscience,botella2018stages,sadler2015wallas}, they differed substantially in how much they invested in divergence and convergence in practice.}
Most participants favored \name{} for its support of open-ended exploration, but a few participants (P4, P6, P12) preferred the baseline’s more direct, goal-oriented workflow. 
These designers typically began with clearer design intentions and prioritized efficiently realizing outcomes, which aligns more with engineering design models \cite{howard2008describing, wallas1926art}.
For them, the emphasis on continuous variation in our tool introduced friction. 
This suggests that designers’ preferences are shaped not only by context but also by their familiarity with the design task and the clarity of their initial ideas.
\rhl{Interaction logs further revealed different preferences for how designers converged: decomposition was used by nearly all participants (13/14; MD{=3 times}), while zooming and anchoring were each used by 9 of 14 participants, with some relying more heavily on zooming (MAX=9 times) and others on anchoring (MAX=5 times).}
\rhl{This points to opportunities for future interactions to adapt the balance between exploration and refinement, as well as different convergent types, to individual needs and preferences. }
Interfaces could allow direct adjustment of variation level (e.g., via a continuous slider), rather than relying solely on iterative zooming.
\rhl{More robust agentic adaptation could infer a designer’s preferred mode of convergence from their interactions and restructure the generated design space accordingly.}
Enabling systems to respond to both contextual factors and individual preferences may lead to more flexible, personalized workflows.

\textbf{Adaptation to imperfect workflows.}
Despite participants generally rating \name{} more positively and preferring the designs they produced with it, there was no significant difference in overall satisfaction with final outcomes (Q3: MD{=6 > 5}; $p{=0.111}$).
This shows that designers adapted to less supportive tools to achieve comparable results. 
In the baseline, some participants developed workaround strategies, such as iteratively adding and removing moodboard materials to steer generation through trial and error, effectively constructing their own scaffolding process.
This was also reflected in participants’ reflections, as P4 noted that existing AI tools are designed from a programmer’s perspective \cite{adobe_express, microsoft_designer, canva_ai}, but designers have ``\textit{gradually learned to work within this paradigm.}"
These findings suggest that while \name{} better aligns with natural design workflows, designers can compensate for misaligned interactions by adapting their behavior at the cost of increased cognitive effort or less structured processes.
This raises an important implication for future work: improving creative tools may not directly lead to better outcomes, but instead may reshape how effort is distributed throughout the design process. 
Future work could examine how different interaction paradigms influence designers’ cognitive strategies---when they choose to adapt to a system versus when they rely on interaction support, and how these trade-offs impact long-term creativity and workflow development.

\textbf{Extending to human–human collaboration.}
Participants also expressed strong interest in using \name{} in collaborative settings. 
While we focus on human–agent interaction with a single designer, several participants (P1, P3, P8) noted that a key part of their workflow is incorporating external feedback from collaborators. 
They saw potential for \name{} by allowing multiple designers to contribute reference materials and ideas, which the agent can reinterpret into diverse design directions. 
P6 highlighted that their professional workflow often involves workshops with clients and stakeholders, and envisioned the axis-centered interaction as a way to broaden the initial design space before narrowing it down based on client or brand constraints. These findings suggest that \name{} could serve as a shared creative workspace to support collective exploration and negotiation of design directions.
Future work could investigate how axis-centered interaction supports human-human collaboration in creative teams, including how multiple inputs can be integrated, compared, and refined, as well as how systems can balance open-ended exploration with constraint-driven convergence in collaborative settings.

\rhl{
\textbf{Limitations and Future Work.}
We used the same topic across conditions to control for differences in design-space constraints, but this may have introduced carryover effects. 
We mitigated this through counterbalancing and by encouraging participants to explore new directions when switching conditions. 
An order-based analysis ($n{=}7$ per group) still showed higher average ratings for \name{} on both the post-questionnaire (baseline-first: M=5.51 > 4.56; S2R-first: M=5.20 > 4.34) and CSI (baseline-first: M=6.85 > 4.95; S2R-first: M=6.32 > 5.52).
However, carryover cannot be fully ruled out and future work could use separate tasks with carefully matched design-space constraints. 
We also used a self-built baseline to isolate the effects of the axis-centered workflow, but future work could complement this controlled comparison by also comparing to existing commercial tools.
Regarding technical limitation, our decomposition (Sec. ~\ref{sec:decompose}) synthesizes reusable concept representations rather than segmenting visual objects. 
Consequently, it may produce overly abstract or entangled concepts for ambiguous designs. 
Future work could explore object-level representations to improve decomposition fidelity and control. 
In terms of generalizability, we focused on creative design and targeted professional designers in this work, but future work could investigate how well the proposed workflow transfers to other creative tasks or more novice designers. 
We see potential for extending the workflow to domains such as interaction design, 3D modeling, and video creation, although each medium introduces distinct dimensions. 
Future research could explore how axis-centered workflows can be adapted to accommodate such domain-specific dimensions while preserving support for structured exploration and refinement.
}
\section{Conclusion}
\shl{We propose an axis-centered workflow for agent-assisted design that adaptively structures the design space across exploration and refinement in creative scaffolding.
We instantiate this workflow in \name{}, a prototype that allows designers to build and reshape a dynamic $n\times$n design space grounded in moodboard materials and interpretable axes.
The system supports broad yet structured exploration through meaningful ``surprises,'' followed by progressively focused refinement through zooming, anchoring, and decomposition.}
We conducted a within-subjects study with 14 designers, comparing \name{} against a baseline in poster design. 
Participants rated \name{} more positively than the baseline, reporting increased sense of control and improved ability to navigate their scaffolding path. 
While both tools achieved comparable satisfaction in final outcomes, designers generally found \name{} to better support their workflows and sensemaking, producing designs perceived as more creative and meaningful.
We discuss design implications for agent-assisted creative systems and outline directions for future research in supporting structured, interactive exploration and refinement.

\begin{acks}
This work is supported in part by the Natural Sciences and Engineering Research Council of Canada (NSERC) Discovery Grant \#RGPIN-2020-03966, the Ontario Early Researcher Award (ERA) \#ER24-18-222, the Canada Foundation for Innovation (CFI) John R. Evans Leaders Fund (JELF) \#42371, and a gift fund from Adobe.
\end{acks}

\bibliographystyle{ACM-Reference-Format}
\bibliography{sample-base}

\appendix
\section{Additional Implementation Details}
\label{implementation_details}

In this appendix, we provide additional implementation details of our system, including the end-to-end pipeline and the prompt templates used by the agents.

\subsection{Initial Scale and Grid Generation}
\label{initial_gen_appendix}

Initial grid generation is implemented as an agent-driven two-stage pipeline. Upon user prompt submission, the agent first triggers scale synthesis by aggregating deduplicated moodboard concepts and default or UI-defined axis semantics, then constructs a structured prompt and queries a VLM (Gemini-2.5-Flash-Lite) to generate two 3-point ordinal scales (each with three lexical variations) in a validated JSON schema. The agent then materializes cell-level constraints by mapping these scales onto the 3×3 grid, assigning each cell a pair of X/Y variations based on its row-column indices. In the second stage, the agent performs nine parallel image generations (Gemini-2.5-Flash-Image), where each cell’s prompt is assembled through a template-ordered concatenation of references, user annotations, and axis-cross-product constraints.

Below, we present the prompt templates used for each step:

\medskip %
\noindent\textbf{Moodboard Concept Extraction Template:}
\begin{lstlisting}[
    basicstyle=\scriptsize\ttfamily,
    breaklines=true,
    postbreak=\mbox{\textcolor{gray}{$\hookrightarrow$}\space},
    frame=single,
    backgroundcolor=\color{gray!3},
    xleftmargin=2pt,
    xrightmargin=2pt,
    showstringspaces=false,
    columns=flexible,
    aboveskip=10pt, %
    belowskip=10pt
]
Analyze this <image|sketch> and extract exactly 20 key visual concepts as keywords. Each concept should be 1-5 words long. Focus on visual elements, style, composition, colors, subjects, mood, techniques, and artistic elements visible in the image.

IMPORTANT: Return ONLY a simple comma-separated list of keywords. 
Do NOT provide descriptions or sentences.

Example:
"vibrant colors, dramatic lighting, modern architecture, stylized characters, clean lines, bold outlines, graphic design, surreal mood, dynamic composition, detailed clothing, crisp shading, contrasting tones, hand gestures, golden sky, flowing hair, perspective angles, minimalist background, character design, visual storytelling, atmospheric lighting"

Analyze the actual <image|sketch> content and extract keywords that accurately describe what you see.
\end{lstlisting}

\medskip
\noindent\textbf{Initial Scale Generation Template:}
\begin{lstlisting}[
    basicstyle=\scriptsize\ttfamily, %
    breaklines=true,
    postbreak=\mbox{\textcolor{gray}{$\hookrightarrow$}\space},
    frame=single,
    backgroundcolor=\color{gray!3},
    xleftmargin=2pt,
    xrightmargin=2pt,
    showstringspaces=false,
    columns=flexible,
    aboveskip=5pt,
    belowskip=5pt
]
EXAMPLE:
- If moodboard has "vintage", "retro" -> "vintage sepia", "retro neon"
- If moodboard has "minimalist", "clean" -> "ultra minimal", "clean geometric"

For each axis, create 3 points that:
1. Are derived from moodboard concepts
2. Are 1-5 words each
3. Show clear progression relevant to userPrompt
4. Use actual terminology from concept list
5. Are actionable and specific

For each point, create 3 specific variations that:
1. Are concrete examples of that point
2. Incorporate userPrompt elements
3. Use moodboard concepts as inspiration
4. Are distinct but clearly belong to the same point
5. Are specific enough to guide creation

If user-replaced cells exist:
6. Align axis points with anchored cell semantics
7. Preserve coherent row/column relationships

Return response in this exact JSON format:
{
  "xAxis": {
    "point1": "specific_concept_based_term",
    "point1Variations": ["ex1", "ex2", "ex3"],
    "point2": "specific_concept_based_term",
    "point2Variations": ["ex1", "ex2", "ex3"],
    "point3": "specific_concept_based_term",
    "point3Variations": ["ex1", "ex2", "ex3"]
  },
  "yAxis": {
    "point1": "specific_concept_based_term",
    "point1Variations": ["ex1", "ex2", "ex3"],
    "point2": "specific_concept_based_term",
    "point2Variations": ["ex1", "ex2", "ex3"],
    "point3": "specific_concept_based_term",
    "point3Variations": ["ex1", "ex2", "ex3"]
  }
}
\end{lstlisting}

\noindent\textbf{Initial Grid Image Generation Template:}
\begin{lstlisting}[
    basicstyle=\scriptsize\ttfamily, %
    breaklines=true,
    postbreak=\mbox{\textcolor{gray}{$\hookrightarrow$}\space},
    frame=single,
    backgroundcolor=\color{gray!3},
    xleftmargin=2pt,
    xrightmargin=2pt,
    showstringspaces=false,
    columns=flexible,
    aboveskip=5pt,
    belowskip=5pt
]
[1] Role line (exactly one):
- Only references: Use the reference images provided for visual style and inspiration.
- Only sketches: Use the user sketches provided for composition and layout.
- Both: Images <ref indices> are reference images. Images <sketch indices> are user sketches.

[2] Optional linked text (per asset):
Image <i>: <linkedText>

[3] Optional standalone text:
Additional information provided by the user: <standaloneText>

[4] Optional grid constraint:
IMPORTANT: This image must specifically follow the combination:
<xAxisFullText> "<xVariation>" combined with
<yAxisFullText> "<yVariation>".
\end{lstlisting}

\subsection{Zooming}

The zoom is implemented as a stateful process over a 3$\times$3 lattice, where each zoom level increments \texttt{currentZoomLevel} and is managed through bounded undo/redo stacks storing grid snapshots, axis labels, and optional cached generation results. Zoom-in relocates the selected cell to its mirror position $(2 - r, 2 - c)$ to serve as the anchor for the next level. 
Surrounding cells are regenerated via a two-stage pipeline: either restored from cache or recomputed by extracting keywords from the zoomed image to synthesize new axis scales and generate variations conditioned on these semantics. 
Zoom-out restores prior states while preserving forward history. Overall, zoom operates as a level-based re-instantiation of the axis-constrained grid around a selected patch.

\medskip %
\noindent\textbf{Zoom Scale Generation Template:}
\begin{lstlisting}[
    basicstyle=\scriptsize\ttfamily, %
    breaklines=true,
    postbreak=\mbox{\textcolor{gray}{$\hookrightarrow$}\space},
    frame=single,
    backgroundcolor=\color{gray!3},
    xleftmargin=2pt,
    xrightmargin=2pt,
    showstringspaces=false,
    columns=flexible,
    aboveskip=5pt,
    belowskip=5pt
]
You are creating a 3-point scale for each axis based on analyzing a specific image that represents the intersection point of the scales.

The image is positioned at grid coordinates (<row>, <col>) in a 3x3 grid, which means it represents the combination of:
- X-axis point <col+1> (<xAxisDefinition>)
- Y-axis point <row+1> (<yAxisDefinition>)

Analyzed image concepts: <comma-separated keywords from vision analysis>

Base prompt: "<userPrompt>"

Create two 3-point scales where the intersection point (X-axis point <col+1> + Y-axis point <row+1>) should match the visual characteristics of the analyzed image. Each point should have 3 variations for diversity.

IMPORTANT: The main point descriptions should be concise (1-3 words, but can be 2-3 words when needed for clarity), while the variations can be longer and more descriptive.

CRITICAL: The image is positioned at (<row>, <col>), so:
- X-axis point <col+1> must match the image's <xAxisDefinition> characteristics
- Y-axis point <row+1> must match the image's <yAxisDefinition> characteristics
- All other points should contrast with the image to create a meaningful scale

X-axis scale (<xAxisDefinition>):
- Point 1: [SHORT 1-3 word description<if col===0 then " that matches the image characteristics" else " that contrasts with the image">] with 3 longer variations
- Point 2: [SHORT 1-3 word description<if col===1 then match else contrast>] with 3 longer variations
- Point 3: [SHORT 1-3 word description<if col===2 then match else contrast>] with 3 longer variations

Y-axis scale (<yAxisDefinition>):
- Point 1: [SHORT 1-3 word description<if row===0 then match else contrast>] with 3 longer variations
- Point 2: [SHORT 1-3 word description<if row===1 then match else contrast>] with 3 longer variations
- Point 3: [SHORT 1-3 word description<if row===2 then match else contrast>] with 3 longer variations

Return ONLY a JSON object with this exact structure:
{
  "xAxis": {
    "point1": "short description",
    "point1Variations": ["detailed variation1", "detailed variation2", "detailed variation3"],
    "point2": "short description",
    "point2Variations": ["detailed variation1", "detailed variation2", "detailed variation3"],
    "point3": "short description",
    "point3Variations": ["detailed variation1", "detailed variation2", "detailed variation3"]
  },
  "yAxis": {
    "point1": "short description",
    "point1Variations": ["detailed variation1", "detailed variation2", "detailed variation3"],
    "point2": "short description",
    "point2Variations": ["detailed variation1", "detailed variation2", "detailed variation3"],
    "point3": "short description",
    "point3Variations": ["detailed variation1", "detailed variation2", "detailed variation3"]
  }
}
\end{lstlisting}

\medskip
\noindent\textbf{Zoom Image Generation Template:}
\begin{lstlisting}[
    basicstyle=\scriptsize\ttfamily,
    breaklines=true,
    postbreak=\mbox{\textcolor{gray}{$\hookrightarrow$}\space},
    frame=single,
    backgroundcolor=\color{gray!3},
    xleftmargin=2pt,
    xrightmargin=2pt,
    showstringspaces=false,
    columns=flexible,
    aboveskip=5pt,
    belowskip=5pt
]
The first image is the base image to work from. The poster should be a variation of this zoomed image.

[Multimodal order]
Image 1: zoomed crop (always)
Images 2..N: optional moodboard references and/or sketches

[Optional - extra moodboard images]
- One reference: Image 2 is an additional reference image from the moodboard for inspiration.
- Multiple references: Images 2, 3, ... are additional reference images from the moodboard for inspiration.
- One sketch (after refs): Image <k> is a user sketch from the moodboard for composition guidance.
- Multiple sketches: Images <k>, ... are user sketches from the moodboard for composition guidance.

[Optional - linked text per asset]
Image <i>: <linkedText>

[Optional - standalone moodboard text]
Additional information provided by the user: <standaloneText>

[Optional - per-cell axis constraint]
IMPORTANT: This image must specifically follow the combination:
<xAxis> "<xVariation>" combined with
<yAxis> "<yVariation>".

[Closing - always]
Create: <userPrompt>
\end{lstlisting}

\subsection{Anchoring}

Anchoring is implemented as a cell-level mechanism that allows users to ``pin'' specific grid positions so they remain fixed during generation and inform subsequent scale synthesis.
A cell becomes anchored either when the user explicitly keeps it (marking it as user-defined and copying it to the moodboard) or when content is dragged from the moodboard into the grid, attaching a persistent badge and flag. 
During generation, the system skips these user-defined cells, regenerating only the remaining positions, while preserving the anchored content in place. 
At the same time, anchored cells are incorporated into the agent’s scale synthesis process: the system extracts their grid coordinates and associated visual concepts, and passes this information to the prompt that constructs axis scales, ensuring that specific rows and columns align semantically with these fixed references. 
As a result, anchoring functions as a binary constraint that both preserves user-selected designs and conditions the structure of the generated grid.

\medskip
\noindent\textbf{Anchor Scale Generation Template:}
\begin{lstlisting}[
    basicstyle=\scriptsize\ttfamily,
    breaklines=true,
    postbreak=\mbox{\textcolor{gray}{$\hookrightarrow$}\space},
    frame=single,
    backgroundcolor=\color{gray!3},
    xleftmargin=2pt,
    xrightmargin=2pt,
    showstringspaces=false,
    columns=flexible,
    aboveskip=5pt,
    belowskip=5pt
]
[Shared opening - always]
<userPrompt>
<xAxis>
<yAxis>
<concepts>

User-anchored cells:
- Cell (row <r>, col <c>): <imageSrc / concepts>
...
These anchored cells must define the corresponding axis levels:
- Columns (0-2) map to X-axis points
- Rows (0-2) map to Y-axis points
Ensure row/column consistency and fuse multiple anchors when needed.

[CRITICAL REQUIREMENTS - always]
1. Derive axis points from moodboard concepts
2. Each point is 1-5 words
3. Points show clear ordinal progression
4. Use concrete, actionable terminology
5. Variations must be distinct but coherent

[Additional requirements - only if anchors exist]
6. Align axis points with anchored cell semantics
7. Preserve row/column consistency with multi-anchor fusion

[Shared tail - always]

EXAMPLE of what we want:
- If moodboard has "vintage", "retro", "nostalgic" -> "vintage sepia", "retro neon", "nostalgic film grain"
- If moodboard has "minimalist", "clean", "modern" -> "ultra minimal", "clean geometric", "modern luxury"
- If moodboard has "fantasy", "magical", "mystical" -> "ethereal fantasy", "dark mystical", "whimsical magic"

For each axis, create exactly 3 points that:
1. Are derived from moodboard concepts
2. Are 1-5 words each
3. Form a clear ordinal progression
4. Use concrete, domain-relevant terminology
5. Are specific enough to guide generation

For each point, create exactly 3 variations that:
1. Are concrete realizations of that point
2. Incorporate elements from the user prompt
3. Reflect moodboard concepts
4. Are distinct but semantically consistent
5. Are directly usable for generation

Return your response in this exact JSON format:
{
  "xAxis": {
    "point1": "...",
    "point1Variations": ["...", "...", "..."],
    "point2": "...",
    "point2Variations": ["...", "...", "..."],
    "point3": "...",
    "point3Variations": ["...", "...", "..."]
  },
  "yAxis": {
    "point1": "...",
    "point1Variations": ["...", "...", "..."],
    "point2": "...",
    "point2Variations": ["...", "...", "..."],
    "point3": "...",
    "point3Variations": ["...", "...", "..."]
  }
}

REMEMBER: Use actual moodboard concepts to construct meaningful, prompt-aligned scales.
\end{lstlisting}

\noindent\textbf{Anchor Image Generation Template:}
\begin{lstlisting}[
    basicstyle=\scriptsize\ttfamily,
    breaklines=true,
    postbreak=\mbox{\textcolor{gray}{$\hookrightarrow$}\space},
    frame=single,
    backgroundcolor=\color{gray!3},
    xleftmargin=2pt,
    xrightmargin=2pt,
    showstringspaces=false,
    columns=flexible,
    aboveskip=5pt,
    belowskip=5pt
]
[Role line]
- Only references:
  Use the reference images provided for visual style and inspiration.
- Only sketches:
  Use the user sketches provided for composition and layout.
- Both:
  Images <ref indices> are reference images (for visual style).
  Images <sketch indices> are user sketches (for composition/layout).

[Optional]
Image <i>: <linkedText>
Additional information provided by the user: <standaloneText>

[Axis constraint]
IMPORTANT: This image must specifically follow the combination:
<xAxis> "<xVariation>" combined with <yAxis> "<yVariation>".

[Task]
Create: <userPrompt>
\end{lstlisting}

\subsection{Decomposition}

Decomposition converts a moodboard or generated image into reusable concept assets through a two-phase, agent-driven pipeline. 
First, the image is encoded as a base64 input and sent to the vision–language endpoint, which extracts a small set of structured abstract factors (e.g., composition, mood, style) as a JSON array of typed descriptions that are immediately displayed in the UI. 
In parallel, a second endpoint generates a synthetic composite image conditioned on these factors, enforcing a fixed 3×2 layout of six circular regions on a transparent background. 
The client then deterministically slices this composite into six patches via uniform grid cropping, downscales them, and inserts them into the moodboard as independent visual assets, optionally augmented with per-circle textual descriptions and a derived color palette glyph. 
This process does not rely on traditional segmentation or geometric detection; instead, it combines structured VLM-based factor extraction with reference-conditioned image generation and fixed-layout post-processing to produce both semantic and visual decompositions of the source image.

\medskip
\noindent\textbf{Decomposition Concept Extraction Prompt:}
\begin{lstlisting}[
    basicstyle=\scriptsize\ttfamily,
    breaklines=true,
    postbreak=\mbox{\textcolor{gray}{$\hookrightarrow$}\space},
    frame=single,
    backgroundcolor=\color{gray!3},
    xleftmargin=2pt,
    xrightmargin=2pt,
    showstringspaces=false,
    columns=flexible,
    aboveskip=5pt,
    belowskip=5pt
]
You are an art analysis expert. Return ONLY a valid JSON array of abstract concept objects. No explanations, no markdown.

Analyze this image and identify its abstract artistic and compositional qualities.
Focus on:
- Composition style (e.g., rule of thirds, centered, dynamic)
- Overall theme or narrative (e.g., nature, urban, fantasy)
- Mood and atmosphere (e.g., calm, dramatic, mysterious)
- Artistic style (e.g., realistic, stylized, abstract)
- Visual rhythm and flow (e.g., flowing, structured, chaotic)
- Lighting approach (e.g., dramatic, soft, natural)

Return ONLY a JSON array with 3-4 of the most important abstract concepts, each as:
{
  "type": "composition" | "theme" | "mood" | "style" | "lighting" | "rhythm",
  "description": "brief description"
}

Example:
[
  {"type": "composition", "description": "rule of thirds with strong diagonal lines"},
  {"type": "mood", "description": "serene and peaceful atmosphere"}
]

[Vision input: source image (base64)]
\end{lstlisting}

\noindent\textbf{Decomposition Visual Token Generation Template:}
\begin{lstlisting}[
    basicstyle=\scriptsize\ttfamily,
    breaklines=true,
    postbreak=\mbox{\textcolor{gray}{$\hookrightarrow$}\space},
    frame=single,
    backgroundcolor=\color{gray!3},
    xleftmargin=2pt,
    xrightmargin=2pt,
    showstringspaces=false,
    columns=flexible,
    aboveskip=5pt,
    belowskip=5pt
]
[Base instruction - always]
Generate an image consisting of exactly 6 circular regions arranged in a 3x2 grid on a transparent background. Each circle should represent a distinct concept derived from the input image.

[Condition: abstract concepts exist]
The image has these abstract qualities: <type: description, ...>.
Create 6 circles:
- 2-3 representing concrete visual elements
- 2-3 representing abstract concepts
Ensure each circle is visually distinct.

[Else: no abstract concepts]
Create 6 circles representing both concrete visual elements and abstract concepts (composition, mood, theme, style). Make each circle unique and visually distinct.

[CRITICAL constraint - always]
CRITICAL: Your output must have exactly 6 circles in a 3x2 grid. No more. If you are about to add a 7th circle, stop. Only 6 circles.

[Reference input]
- Original user image (as visual reference)
- Prompt above (model may append resolution instructions)
\end{lstlisting}

\noindent\textbf{Decomposition Color Palette Extraction Template:}
\begin{lstlisting}[
    basicstyle=\scriptsize\ttfamily,
    breaklines=true,
    postbreak=\mbox{\textcolor{gray}{$\hookrightarrow$}\space},
    frame=single,
    backgroundcolor=\color{gray!3},
    xleftmargin=2pt,
    xrightmargin=2pt,
    showstringspaces=false,
    columns=flexible,
    aboveskip=5pt,
    belowskip=5pt
]
You are a color analysis expert. You MUST return ONLY a valid JSON array of color objects. No explanations, no markdown.

Analyze this image and identify the 6-8 most important colors.
Consider:
- Dominant colors defining mood and atmosphere
- Key accent colors that draw attention
- Colors representing important visual elements
- Colors contributing to artistic style

Return ONLY a JSON array in this exact format:
[
  {"r": 255, "g": 200, "b": 100, "description": "warm golden yellow"},
  {"r": 50, "g": 100, "b": 150, "description": "deep ocean blue"}
]

Each color must include:
- r, g, b: integer RGB values (0-255)
- description: brief semantic description

Return ONLY the JSON array, no other text.
\end{lstlisting}

\begin{figure*}[t]
    \centering
    \includegraphics[width=\linewidth]{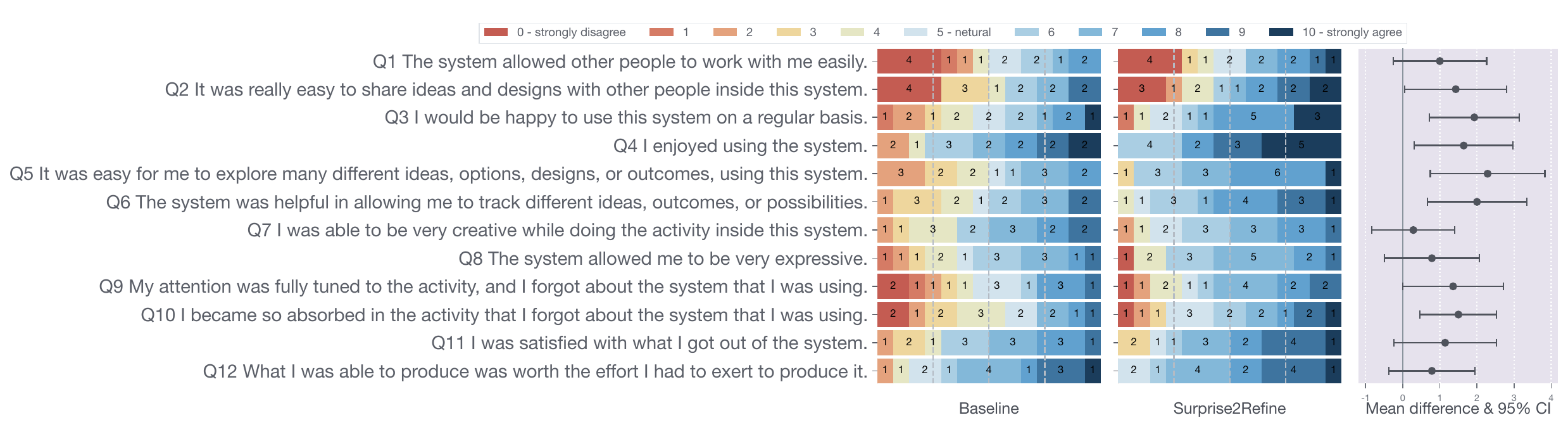}
    \caption{Participants’ CSI questionnaire responses for the baseline and \name{}. Dots indicate mean differences between the two conditions.}
    \Description{A chart showing participants’ responses to 12 Creativity Support Index (CSI) questionnaire items for two conditions: Baseline and Surprise2Refine. Each row corresponds to a question about collaboration, creativity, expressiveness, engagement, and satisfaction. For each condition, stacked horizontal bars display response distributions on a 10-point Likert scale from “strongly disagree” (0) to “strongly agree” (10). On the right, a dot-and-error-bar plot shows the mean difference between conditions with 95 percent confidence intervals for each question. Across all items, Surprise2Refine shows higher agreement scores than the baseline, indicating improved support for creativity, exploration, and user engagement.}
    \label{fig:csi_all}
\end{figure*}  

\section{Additional Evaluation Study Results}
\label{additional_results}

In this appendix, we provide additional quantitative results from our evaluation study.

\subsection{Mean LPIPS Diversity Scores}

\begin{figure*}[bp]
    \centering
    \includegraphics[width=\linewidth]{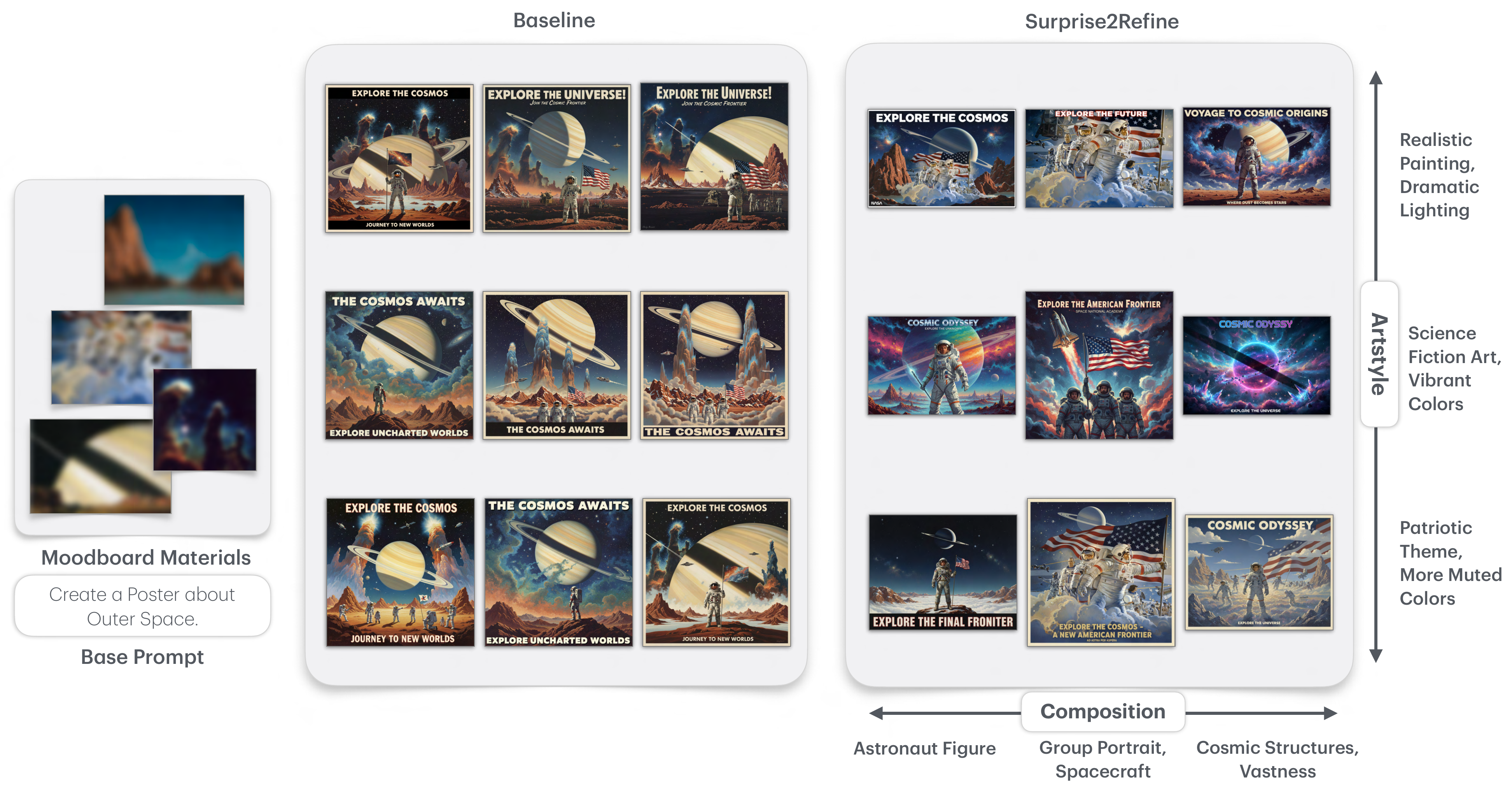}
    \caption{Initial generations in \name{} (Mean LPIPS=0.575, SD=0.047) versus the baseline (Mean LPIPS=0.527, SD=0.055) for P1.}
    \Description{A side-by-side comparison of initial poster generations for an outer space theme from P1. On the left (Baseline), a moodboard and prompt (“Create a Poster about Outer Space”) produce a 3×3 grid of visually similar posters featuring planets, astronauts, and space landscapes with limited variation. On the right (Surprise2Refine), a corresponding set of generated posters shows greater diversity, organized along two axes: Composition (e.g., astronaut figure, group portrait with spacecraft, cosmic structures) and Artstyle (e.g., realistic painting with dramatic lighting, science fiction art with vibrant colors, patriotic theme with more muted colors). The Surprise2Refine results demonstrate broader variation in layout, color, and thematic interpretation compared to the baseline.}
    \label{fig:p1}
\end{figure*}

\begin{figure*}[bp]
    \centering
    \includegraphics[width=\linewidth]{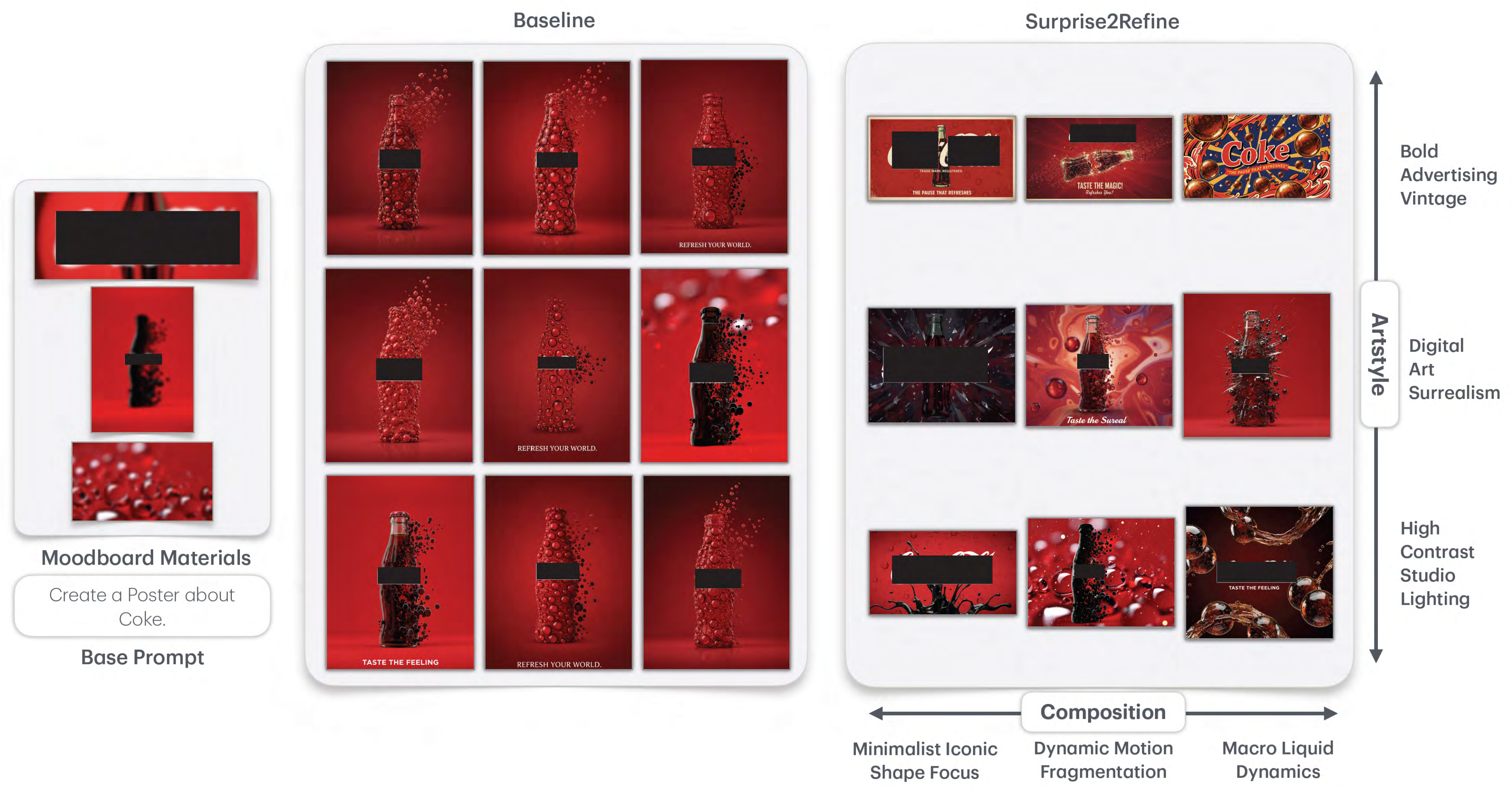}
    \caption{Initial generations in \name{} (Mean LPIPS=0.673, SD=0.080) versus the baseline (Mean LPIPS=0.342, SD=0.135) for P12.}
    \Description{
    A side-by-side comparison of initial poster generations for a Coke theme from P12. On the left (Baseline), a moodboard and prompt (“Create a Poster about Coke”) produce a 3×3 grid of highly similar posters featuring a centered Coca-Cola bottle on a red background with minimal variation. On the right (Surprise2Refine), generated posters show greater diversity, organized along two axes: Composition (e.g., minimalist iconic shape focus, dynamic motion fragmentation, macro liquid dynamics) and Artstyle (e.g., bold advertising vintage, digital art surrealism, high contrast studio lighting). The Surprise2Refine results demonstrate broader variation in layout, visual effects, and stylistic interpretation compared to the baseline.
    }
    \label{fig:p12}
\end{figure*}

In Table~\ref{tab:lpips_diversity}, we report the average LPIPS diversity scores across all participants for the initial nine generated designs in \name{} and the baseline, computed using the same set of moodboard references and input prompts. 
Our results show that, in the early exploration stage, the generations produced by \name{} exhibit higher diversity than those from the baseline across all participants.
Below, Figures~\ref{fig:p1}$\sim$\ref{fig:p12} showcase example initial generations in the exploration stage from \name{} and the baseline for P1, P3, P8, and P12. 
The input images have been redacted for copyright reasons.

\begin{table}[H] 
\centering
\small %
\setlength{\tabcolsep}{6pt} %
\begin{tabular}{lcc}
\toprule
\textbf{Participant} & \textbf{Baseline: Mean (SD)} & \textbf{\name{}: Mean (SD)} \\
\midrule
P1  & 0.527 (0.055) & 0.575 (0.047) \\
P2  & 0.484 (0.045) & 0.561 (0.124) \\
P3  & 0.618 (0.051) & 0.729 (0.041) \\
P4  & 0.571 (0.093) & 0.651 (0.051) \\
P5  & 0.583 (0.052) & 0.713 (0.080) \\
P6  & 0.636 (0.074) & 0.736 (0.088) \\
P7  & 0.551 (0.083) & 0.650 (0.054) \\
P8  & 0.583 (\rhl{0.102}) & 0.612 (0.086) \\
P9  & 0.495 (0.065) & 0.617 (0.098) \\
P10 & 0.453 (0.103) & 0.616 (\rhl{0.136}) \\
P11 & 0.537 (0.081) & 0.621 (0.083) \\
P12 & 0.342 (\rhl{0.135}) & 0.673 (0.080) \\
P13 & 0.509 (0.041) & 0.656 (0.061) \\
P14 & 0.522 (\rhl{0.110}) & 0.568 (0.089) \\
\midrule
\textbf{Overall} & \textbf{0.529 (0.078)} & \textbf{0.641 (0.080)} \\
\bottomrule
\end{tabular}
\caption{Mean LPIPS diversity across participants.}
\Description{A table reporting mean LPIPS diversity scores (with standard deviation in parentheses) for 14 participants (P1 to P14) under two conditions: Baseline and Surprise2Refine. Each row lists a participant’s mean and SD for both conditions. Across all participants, Surprise2Refine shows higher mean LPIPS values than the baseline, indicating greater diversity. The final row summarizes overall results: Baseline mean 0.529 (SD 0.080) and Surprise2Refine mean 0.641 (SD 0.080), showing an overall increase in diversity with Surprise2Refine.}
\label{tab:lpips_diversity}
\vspace{-30pt}

\end{table}

\begin{table*}[tbp]
\centering
\small
\begin{tabular*}{\textwidth}{@{\extracolsep{\fill}}lcccccc}
\toprule
& & \multicolumn{2}{c}{\textbf{Avg. Factor Score (SD)}} & \multicolumn{2}{c}{\textbf{Avg. Weighted Score (SD)}} \\
\cmidrule(lr){3-4} \cmidrule(lr){5-6}
\textbf{Scale} & \textbf{Avg. Counts} & \textbf{Baseline} & \textbf{\name{}} & \textbf{Baseline} & \textbf{\name{}} \\
\midrule
Results Worth Effort & 3.43 (1.34) & 12.79 (3.60) & 14.71 (3.41) & 42.86 (20.41) & 50.93 (24.94) \\
Exploration          & 3.14 (1.23) & 10.21 (3.95) & 14.50 (2.71) & 32.71 (20.19) & 45.43 (18.58) \\
Collaboration        & 0.57 (1.40) & 7.79 (5.75)  & 10.21 (7.08) & 8.43 (21.24)  & 9.57 (25.08)  \\
Enjoyment            & 2.36 (1.28) & 11.79 (4.92) & 15.36 (3.61) & 27.14 (20.07) & 35.21 (21.20) \\
Expressiveness       & 3.14 (1.35) & 11.36 (4.48) & 12.43 (3.78) & 33.93 (17.56) & 37.57 (19.88) \\
Immersion            & 2.36 (1.34) & 8.93 (5.64)  & 11.79 (5.28) & 22.29 (19.04) & 30.43 (23.15) \\
\bottomrule
\end{tabular*}
\caption{CSI results comparing baseline and \name{}.}
\Description{A table summarizing Creativity Support Index (CSI) results across six factors: Results Worth Effort, Exploration, Collaboration, Enjoyment, Expressiveness, and Immersion. For each factor, the table reports average counts, average factor scores (with standard deviation), and average weighted scores (with standard deviation) for both Baseline and Surprise2Refine conditions. Across all factors, Surprise2Refine shows higher average factor scores and weighted scores than the baseline, with particularly notable improvements in Exploration, Enjoyment, and Results Worth Effort, indicating stronger support for creative engagement and user experience.}
\vspace{-20pt}
\label{tab:csi}
\end{table*}

\subsection{CSI Results}

In Figure~\ref{fig:csi_all}, we present all participants’ CSI ratings for \name{} and the baseline. 
Overall, \name{} was rated more positively across all items. 
Table~\ref{tab:csi} further reports CSI scores for each scale, including results worth effort, exploration, collaboration, enjoyment, expressiveness, and immersion. 
Across all scales, our tool consistently achieves higher scores than the baseline.

\subsection{Design Galleries}

\begin{figure*}[bp]
    \centering
    \includegraphics[width=0.9\linewidth]{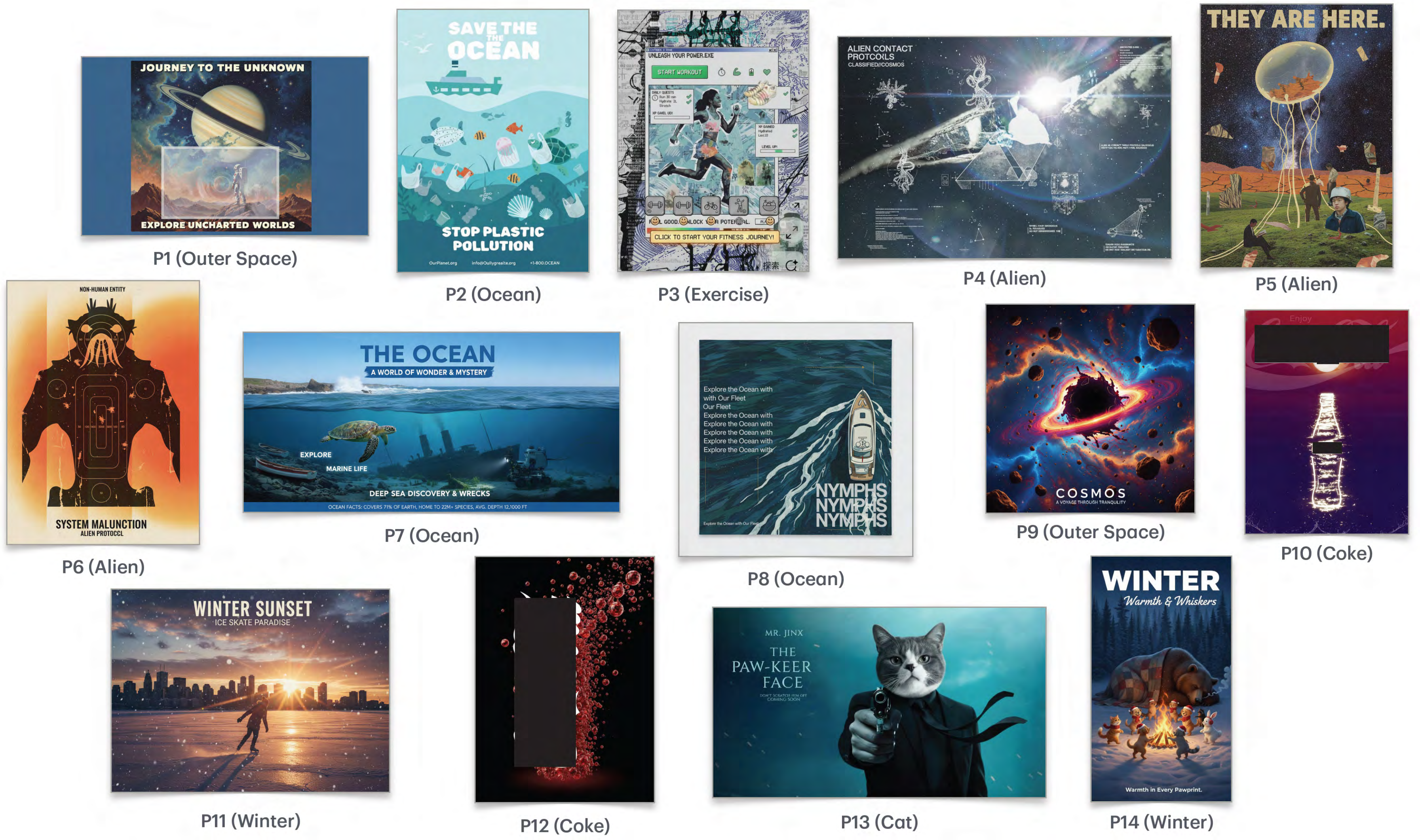}
    \caption{Final designs created by participants (P1$\sim$P14) with the baseline.}
    \Description{
    A collage of final poster designs created by 14 participants (P1 to P14) using the baseline tool. Each poster is labeled with a participant ID and theme. The designs span a range of topics, including outer space (e.g., planets and astronauts), ocean (e.g., marine life and underwater scenes), exercise (e.g., fitness interfaces), alien themes (e.g., sci-fi figures and environments), Coca-Cola branding (e.g., stylized bottles and logos), winter scenes (e.g., snowy landscapes and activities), and a cat-themed poster.
    }
    \label{fig:baseline_design}
\end{figure*}

\begin{figure*}[bp]
    \centering
    \includegraphics[width=0.9\linewidth]{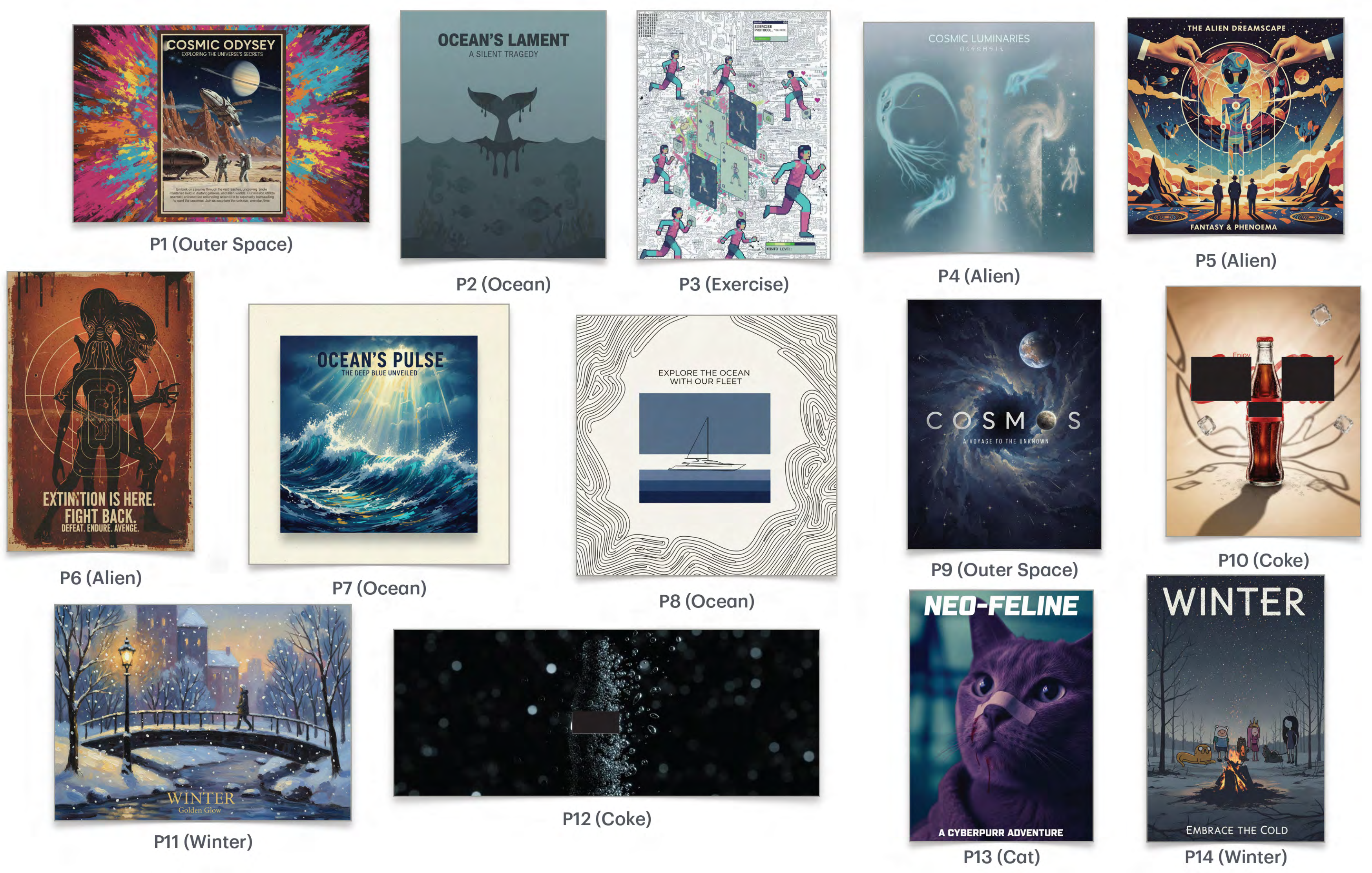}
    \caption{Final designs created by participants (P1$\sim$P14) with \name{}.}
    \Description{
    A collage of final poster designs created by 14 participants (P1 to P14) using Surprise2Refine. Each poster is labeled with a participant ID and theme. The designs span a range of topics, including outer space (e.g., planets and astronauts), ocean (e.g., marine life and underwater scenes), exercise (e.g., fitness interfaces), alien themes (e.g., sci-fi figures and environments), Coca-Cola branding (e.g., stylized bottles and logos), winter scenes (e.g., snowy landscapes and activities), and a cat-themed poster.
    }
    \label{fig:surprise2refine_design}
\end{figure*}

We present the final designs created by participants using the baseline (Fig.\ref{fig:baseline_design}) and \name{} (Fig.\ref{fig:surprise2refine_design}) at the end of this appendix. 

\begin{figure*}[tbp]
    \centering
    \includegraphics[width=\linewidth]{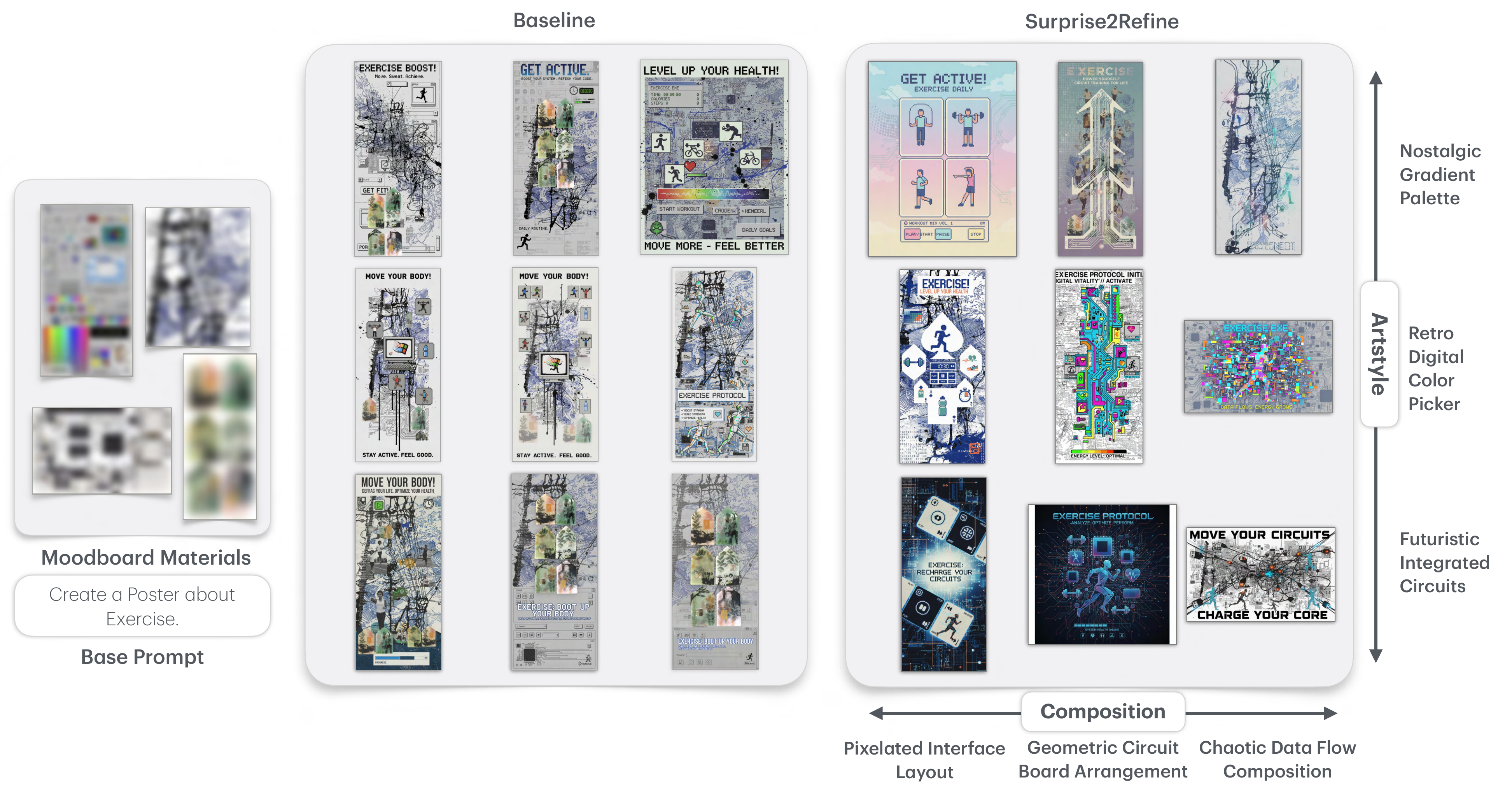}
    \caption{Initial generations in \name{} (Mean LPIPS=0.729, SD=0.041) versus the baseline (Mean LPIPS=0.618, SD=0.051) for P3.}
    \Description{A side-by-side comparison of initial poster generations for an exercise theme from P3. On the left (Baseline), a moodboard and prompt (“Create a Poster about Exercise”) produce a set of visually similar posters with limited variation in layout and style. On the right (Surprise2Refine), generated posters show greater diversity, organized along two axes: Composition (e.g., pixelated interface layout, geometric circuit board arrangement, chaotic data flow composition) and Artstyle (e.g., nostalgic gradient palette, retro digital color picker, futuristic integrated circuits). The Surprise2Refine results demonstrate broader variation in structure, visual style, and interpretation compared to the baseline.}
    \label{fig:p3}
\end{figure*}  

\begin{figure*}[tbp]
    \centering
    \includegraphics[width=\linewidth]{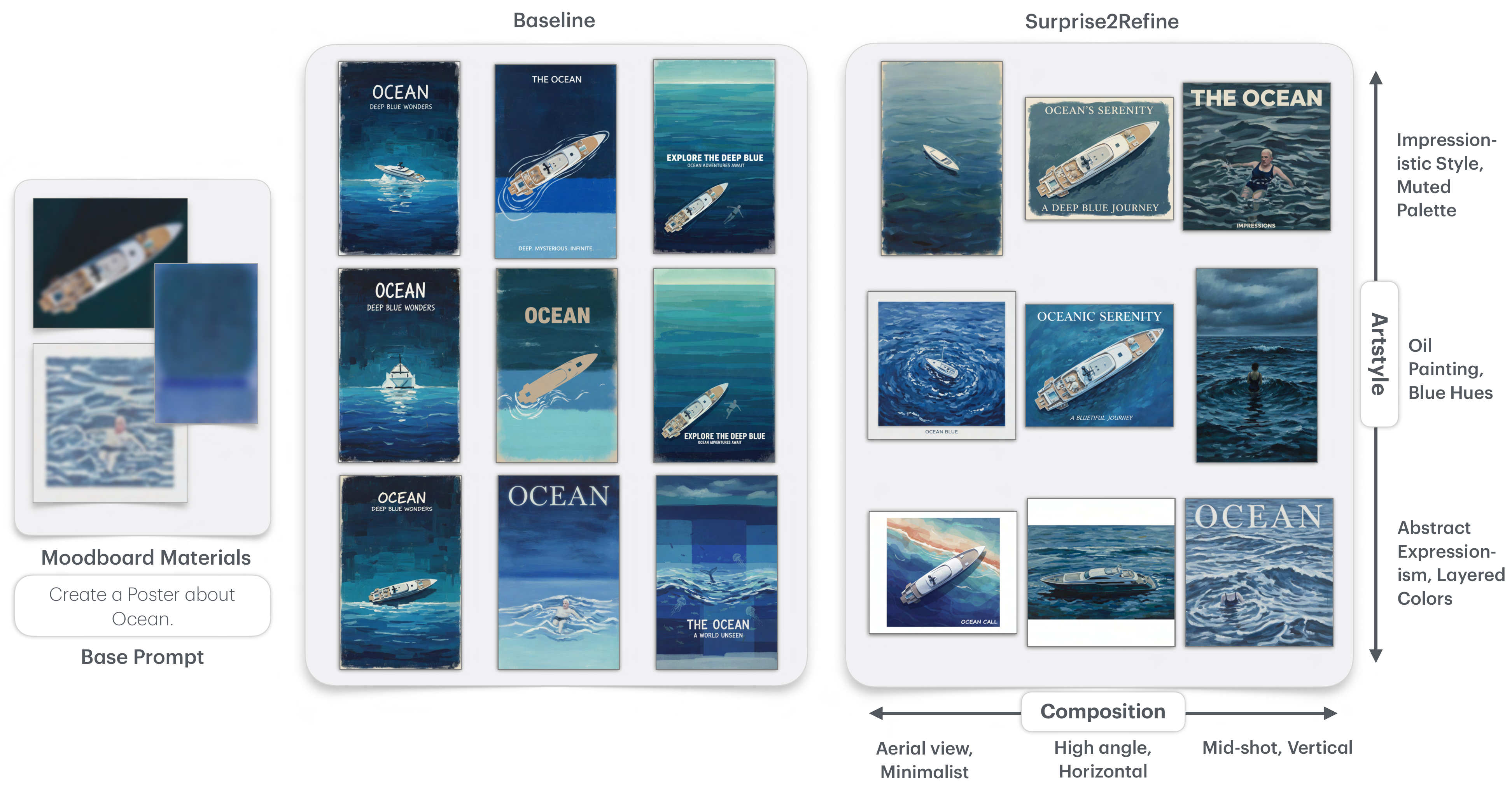}
    \caption{Initial generations in \name{} (Mean LPIPS=0.612, SD=0.086) versus the baseline (Mean LPIPS=0.583, SD=0.102) for P8.}
    \Description{
    A side-by-side comparison of initial poster generations for an ocean theme from P8. On the left (Baseline), a moodboard and prompt (“Create a Poster about Ocean”) produce a 3×3 grid of visually similar posters featuring boats and ocean scenes with limited variation. On the right (Surprise2Refine), generated posters show greater diversity, organized along two axes: Composition (e.g., aerial view, minimalist; high angle, horizontal; mid-shot, vertical) and Artstyle (e.g., impressionistic style with muted palette; oil painting with blue hues; abstract expressionism with layered colors). The Surprise2Refine results demonstrate broader variation in layout, perspective, and visual style compared to the baseline.
    }
    \label{fig:p8}
\end{figure*}  

\end{document}